\documentclass[useAMS, usenatbib]{mn2e}
\usepackage{amsmath}
\usepackage{amssymb}
\usepackage{graphicx}  
\usepackage{pdflscape}

\DeclareMathOperator{\sech}{sech}   
\def\msun{\,M_\odot}
\def\pc{\,{\rm pc}}
\def\kpc{\,{\rm kpc}}
\def\kms{\,{\rm km\,s^{-1}}}
\def\myr{\,{\rm Myr}}
\def\gyr{\,{\rm Gyr}}
\def\feh{\left[{\rm Fe/H}\right]}
\def\e{{\rm e}}

\title[Heating histories in disc galaxies]{Age velocity dispersion relations and 
heating histories in disc galaxies}
\author[M. Aumer, J. Binney and R. Sch\"onrich]
{Michael Aumer \thanks{E-mail:Michael.Aumer@physics.ox.ac.uk (MA)}, 
James Binney and Ralph Sch\"onrich\\
Rudolf Peierls Centre for Theoretical Physics, 1 Keble Road, Oxford, OX1 3NP, UK}

\date{Accepted 2016 July 4. Received 2016 June 13; in original form 2016 May 2}
\pubyear{2016}
\begin{document}
\label{firstpage}
\pagerange{\pageref{firstpage}--\pageref{lastpage}} 
\maketitle

\begin{abstract}
We analyse the heating of stellar discs by non axisymmetric structures and giant
molecular clouds (GMCs) in $N$-body simulations of growing disc galaxies. The
analysis resolves long-standing discrepancies between models and data by
demonstrating the importance of distinguishing between measured age-velocity
dispersion relations (AVRs) and the heating histories of the stars that make
up the AVR.  We fit both AVRs and heating histories with formulae $\propto
t^\beta$ and determine the exponents $\beta_R$ and $\beta_z$ derived from 
in-plane and vertical AVRs and $\tilde{\beta}_R$ and $\tilde{\beta}_z$ from 
heating histories. Values of $\beta_z$ are in almost all simulations larger
than values of $\tilde{\beta}_z$, wheras values of $\beta_R$ are similar to
or mildly larger than values of $\tilde{\beta}_R$. Moreover, values of 
$\beta_z$ ($\tilde{\beta}_z$) are generally larger than values of 
$\beta_R$ ($\tilde{\beta}_R$). The dominant cause of these relations is the 
decline over the life of the disc in importance of GMCs as
heating agents relative to spiral structure and the bar. We examine how age
errors and biases in solar neighbourhood surveys influence the measured AVR:
they tend to decrease $\beta$ values by smearing out ages and thus measured
dispersions. We compare AVRs and velocity ellipsoid shapes
$\sigma_z/\sigma_R$ from simulations to Solar-neighbourhood data. We conclude
that for the expected disc mass and dark halo structure, combined GMC and
spiral/bar heating can explain the AVR of the Galactic thin disc. Strong
departures of the disc mass or the dark halo structure from expectation spoil
fits to the data.
\end{abstract}


\section{Introduction}

When the stars in the Solar neighbourhood (Snhd) are binned by age, the
velocity dispersion of each bin increases with its age. This age-velocity
dispersion relation (AVR) has been known and studied for decades (e.g.
\citealp{stromberg,parenago, wielen}) and similar relations have now also
been inferred for external galactic discs (\citealp{beasley} for M33,
\citealp{dorman} for M31). It is generally agreed that  understanding the
physics that establishes the  AVR would be a significant step towards
understanding how galaxies form and evolve.

Despite many efforts, the shape of the AVR is still not adequately constrained.
The major constraints on the AVR come from observations of stars in the Snhd.
The measured ages of stars suffer from substantial errors, and samples of
stars with measured ages typically have a selection function that favours
young stars over old \citep{nordstroem}. Alternatively modelling the velocity
dispersions as functions of stellar colour has been used to determine the
shape of the AVR \citep[hereafter AB09]{ab09}.

Whereas the vertical density profile of the Milky Way (MW) is well fitted by
a double-exponential in $|z|$, the AVR is typically described as a simple
power-law in age $\sigma(\tau)\propto \tau^{\beta}$ with exponent $\beta$.
\cite{quillen} claimed to detect a jump in the vertical AVR for ages $\tau>
10\gyr$ and connected this jump to the double-exponential nature of the
density profile.  However, more recent studies find that the AVR in the Snhd
can be reasonably described by a single power-law (\citealp{holmberg}, AB09).
Moreover, \citet{sb09} demonstrated that the observed vertical density
profile is fitted well by a model in which the histories of star formation and disc heating are
continuous, and \cite{bovy12} showed that subsets of stars selected at
different points in the $([\alpha/\hbox{Fe}],[\hbox{Fe/H}])$ abundance plane have scale heights that
vary continuously, with no evidence for a dichotomy.

To constrain the heating processes responsible for the Snhd AVR, three
diagnostics have been used in addition to the value of $\beta$: (i) the axis
ratios $\sigma_z/\sigma_R$ of the velocity ellipsoids of old and young
populations; (ii) the magnitude $\sigma_{{\rm old}}$ of $\sigma_z$ for the
oldest populations; (iii) the vertical density profile of the MW's disc.

Several physical processes work together to establish the AVR. The goals
of this paper are to deepen our understanding of the collaboration of secular
heating processes and to show how the AVR responds to one or the other process
playing a more prominent role. External heating due to interaction with dark
substructure is not considered in this paper.

\citet{spitzer} showed that scattering of stars off small-scale
irregularities in the potential of the Galactic disc would transform the
nearly circular orbits of young stars, which reflect their origin from the
gas disc, into orbits with higher radial and vertical energy. The discovery
of Giant molecular clouds (GMCs) with masses $M_{\rm GMC}\sim 10^{5-7}\msun$
provided a suitable candidate for the heating agent.  Using analytic
arguments, \citet{lacey} showed that GMC heating could have contributed
significantly to the observed AVR, but that the dispersion $\sigma_{{\rm old}}$ of
the oldest stars could be reproduced only if the masses and / or number
density of GMCs was significantly higher in the past than they are now.
Lacey further derived values $\sigma_z/\sigma_R\sim 0.8$ and $\beta\sim 0.25$
that are, respectively, larger and smaller than the data indicate.
\citet{ida} used analytical calculations more sophisticated than that of
Lacey to show that scattering by GMCs in discs actually yields values
$\sigma_z/\sigma_R\sim 0.5-0.6$ that are consistent with observations.  This
was confirmed with particle simulations by \citet{shiidsuka} and
\citet{haenninen}: to obtain correct values it is essential to take into
account that in a thin disc impact parameters are concentrated towards the
galactic plane, whereas Lacey had assumed an isotropic distribution of impact
parameters \citep{sellwood}.  However, the conflict between observations and
Lacey's values for $\beta$ and $\sigma_{{\rm old}}$ remained
\citep{haenninen}.

GMCs are not the only sources of a non-axisymmetric component to the
gravitational field experienced by disc stars, and any such component will
heat the disc.  \citet{barbanis} showed that spiral structure could 
significantly heat the disc, when the spiral pattern has either a very high density contrast
or is of a transient and recurring nature.  Two-dimensional simulations of
discs continuously fed with young, cold particles showed that transient and
recurrent spiral structure is always present in star forming discs and
provides sufficient heating to explain the in-plane AVR \citep{selcar,
carsel}. From such two-dimensional simulations and analytical
arguments for vertical cloud heating, \citet{carlberg} concluded
that a combination of GMC and spiral heating could explain the observations.
However, spirals do not directly increase $\sigma_z$ significantly (e.g.\
\citealp{sellwood13, martinez}), and the question remained open whether deflections by
GMCs can convert in-plane heat to vertical heat in an appropriate manner.
\citet{jenkins} used analytic arguments to examine this question for growing
discs. They concluded that the observed value of $\sigma_z/\sigma_R$ could be
explained, but the observed value of $\sigma_{{\rm old}}$ was problematic.

Another continuous secular heating process that has been discussed in the literature
is heating by the bar \citep{saha, grand}. The interaction of galactic discs with satellite
galaxies and the corresponding dark matter substructure can also cause disc heating \citep{velazquez}. 
However, in the case of the MW this process has likely only made a minor contribution to the observed AVRs
\citep{just}.

Another contributor to the AVR could be a decline over cosmic time in the
velocity dispersion of stars at their time of birth as discs have become less
gas-rich and and less turbulent \citep{bournaud, forbes}. These models are
motivated by observations of gas kinematics in disc galaxies at various redshifts,
mostly based on H$\alpha$ emission. These observations have 
revealed significantly ($\sim3-10$ times) higher velocity dispersions $\sigma$
at redshifts $z\sim 2-4$ than in corresponding observations of the local universe (e.g. 
\citealp{sins}), and a decline of $\sigma$ with decreasing redshift \citep{wisn}.
It is, however, unclear how the kinematics of young stars which form from 
cold gas, relate to these observations.

Fully cosmological hydrodynamical simulations of galaxy formation have recently 
reached reasonable levels of success in reproducing MW like disc galaxies and 
the AVRs in some of these simulations have been studied \citep{house, bird, martig, 
grand}. At $z=0$ the stellar populations in the majority of these simulations 
are significantly hotter than the stars in the MW at all ages (but see model 
\emph{g92} of Martig et al.). Especially young stars have overly high $\sigma_z$ 
which has been linked to numerical effects and insufficient resolution and shown
to depend on the specifics of the numerical sub-grid star-formation  models 
\citep{house, martig}. Better agreement with the Snhd AVR is generally found 
for galaxies unaffected by mergers. Martig et al. find that the thin disc stars 
in their more successful models are born cold and heated with time, which they
attribute to heating by spirals and overdensities and possibly the coupling between
transient spirals and weak bending waves \citep{masset}. Note that GMCs are not 
resolved in these simulations.

For many years it was assumed that the chemodynamical evolution of any
annulus of the Galactic disc could be modelled in isolation of other annuli.
Now there is clear evidence that radial migration of stars within discs is an
important process \citep{sellwoodb,roskar, sb09, kordopatis}, with the
consequence that the production of a hot population in one annulus, for
example through the action of a bar, can subsequently endow a distant annulus
with a hot population that could not have been locally heated. On account of
radial migration, it is essential to understand the origin of the AVR
\emph{globally}, that is by tracking the evolution of the disc at all radii.
In general we expect the mean birth radius of a coeval cohort of Snhd stars
will decrease with increasing age, and on account of the radial gradient in
velocity dispersion the decrease in birth radius will be reflected in the AVR
\citep{sb09}.

In this paper we use the simulations presented in \citet[hereafter
ABS16]{abs16} to study the formation of the AVR.  Unlike the previously cited
studies, these simulations include simultaneously all the following important
aspects: growing discs with multiple coeval populations, GMCs, recurring
spiral structure with evolving properties, a bar, an evolving GMC-to-stellar
mass fraction, radial migration and sufficiently cold young stars. ABS16 showed 
that although the vertical
profiles of their models do not show a thick disc like that of the MW, some
models do provide quite good fits to the AVR of the Snhd. Hence they
concluded that the thick disc requires additional sources of heat, but the
thin disc can be explained by combined GMC and spiral heating. They showed
that the efficiency of GMC heating declines over time because the fraction of
the disc's mass contained in GMCs falls steadily as a consequence of a
declining star-formation rate (SFR) and a growing disc mass. Their
simulations are thus a promising tool to study what shapes the AVR in thin
galactic discs.

Two major conclusions will emerge from our study: (i) biased ages and age
uncertainties cause measured AVRs to deviate significantly from the true
AVRs; (ii) it is vital to distinguish between an AVR $\sigma(\tau)$, which
gives velocity dispersion as a function of age for stars that are now
co-located, and a \emph{heating history} $\sigma(t-t_{\rm b})$, which gives the
velocity dispersion as a function of time for a cohort of currently
co-located stars that were born at a given time $t_{\rm b}$. Whereas the AVR
$\sigma(\tau)$ for $\tau\simeq4.5\gyr$ quantifies the current kinematics of stars
born contemporaneously with the Sun, the heating history $\sigma(t-t_{\rm b})$ for $t-t_{\rm b}\simeq4.5\gyr$
quantifies the kinematics of stars $4.5\gyr$ after they were born, which
would be $5.5\gyr$ ago in the case of $10\gyr$ old stars in the disc. If stars
were born into a statistically stationary environment providing heating 
processes which are constant in time, the cohort born
$10\gyr$ ago would $5.5\gyr$ ago have been in the same dynamical state that the
Sun's cohort is in now. That is, given a stationary environment the AVR would be
the same function of $\tau$ that the heating history is of $t-t_{\rm b}$.
If a galaxy undergoes a major merger, stars born before and
after the merger will undergo different heating histories $\sigma(t-t_{\rm b})$.
Here we argue, that even in the absence of mergers or declining birth dispersions,
the thin discs of galaxies change beyond recognition over cosmological timescales,
so the environment is very far from stationary, and the heating
experienced by stars born $10\gyr$ ago during the first Gyr of their lives was
very different from the environment experienced by recently born stars during 
the first Gyr of their lives. Consequently heating histories are described 
by entirely different functions from the AVR.

Nevertheless we will find that both AVRs and
heating histories can be well approximated by the modified power law
\begin{equation}
\sigma(x)=\sigma_{10}\left({{x + x_1} \over {10\gyr + x_1}}\right)^{\beta}
\label{eq:heatlaw}
\end{equation}
used by AB09, with $x=\tau$ or $t$. To differentiate between parameters derived from AVRs 
and heating histories, we will mark the latter with a tilde, i.e. $\tilde{\beta}$, 
$\tilde{\sigma}_{10}$ etc. We will find that the indices $\beta$ and $\tilde{\beta}$ 
of these power laws are often dissimilar. Moreover, we find that in the case of a
heating history the value of $\tilde{\sigma}_{10}$ can evolve strongly with the 
time $t_{\rm b}$ of the cohort's birth, whereas in most models $\tilde{\beta}$ evolves only mildly.

Our paper is organised as follows: In Section \ref{sec:simulations} we
briefly describe the simulations.  In Section \ref{sec:biases} we examine the
effects of observational age errors and biases on the AVR.  In Section
\ref{sec:AVR} we describe the model AVRs and compare them to local data.
Topics discussed include the uncertainties of the comparisons arising from
azimuthal variations in the model AVRs (Sections \ref{sec:azimuth} and
\ref{sec:specifics}), and the
diagnostic content of the axis ratios of velocity ellipsoids (Section
\ref{sec:arat}), and power-law fits to AVRs (Section \ref{sec:heatIndex}).
In Section \ref{sec:heathistory} we consider the heating histories for
different populations of coeval  model stars and show how these relate to
AVRs.  In Section~\ref{sec:discuss} we relate our findings to the physics of
star scattering, and we conclude in Section \ref{sec:conclude}.

\tabcolsep=4.5pt
\begin{table*}
\vspace{-0cm}
  \caption{List of models analysed in this paper. 
           {\it 1st Column}: Model Name;
           {\it 2nd Column}: Initial Conditions;
           {\it 3rd Column}: Total baryonic IC mass $M_{\rm{b,i}}$;
           {\it 4th Column}: IC DM halo scalelength {$a_{\rm halo}$};
           {\it 5th Column}: IC radial disc scalelength $h_{R,{\rm disc}}$;
           {\it 6th Column}: IC vertical disc scaleheight $z_{0,{\rm disc}}$;
           {\it 7th Column}: GMCs Yes/No;
           {\it 8th Column}: Cutoff: Adaptive (no new particles in bar region) or 
           fixed (pre-defined evolving inner cutoff for new particles);
           {\it 9th Column}: Total inserted baryonic model mass $M_{\rm f}$ 
           (including initial baryonic mass);
           {\it 10th Column}: Initial disc scalelength $h_{R, {\rm i}}$;
           {\it 11th Column}: Final disc scalelength $h_{R, {\rm f}}$;
           {\it 12th Column}: Scalelength growth parameter $\xi$;
           {\it 13th Column}: Exponential decay timescale $t_{\rm SFR}$ 
           for the star formation rate;
           {\it 14th Column}: Initial velocity dispersion for inserted stellar 
           particles, {$\sigma_0$};
           {\it 15th Column}: GMC star formation efficiency $\zeta$;
}
  \begin{tabular}{@{}ccccccccccccccc@{}}\hline
1st   & 2nd   & 3rd            & 4th         & 5th             & 6th             &7th   & 8th    & 9th             & 10th           & 11th           & 12th    &  13th        & 14th       & 15th  \\
{Name}&{ICs}  &{$M_{\rm{b,i}}$}  &{$a_{\rm halo}$}&{$h_{R,{\rm disc}}$}&{$z_{0,{\rm disc}}$}&{GMCs}&{Cutoff}&{$M_{\rm f}/\msun$}&{$h_{R, {\rm i}}$}& {$h_{R, {\rm f}}$}& {$\xi$} & {$t_{\rm SFR}$}&{$\sigma_0$}& {$\zeta$}\\ 
      &       &{$[10^{9}\msun]$}&{$\kpc$}     &{$\kpc$}         &{$\kpc$}         &      &        &{$[10^{10}]$}     &{$\kpc$}        & {$\kpc$}       &	       & {$[\rm Gyr]$}&{$[\kms]$}  &\\ \hline
   
Y1	    & Y  & 5           & 30.2        & 1.5             & 0.1             & Yes  & Adap   & 5               & 1.5            & 4.3           & 0.5     & 8.0          & 6         & 0.08   \\ 
Y1s2        & Y  & 5           & 30.2        & 1.5             & 0.1             & Yes  & Adap   & 5               & 1.5            & 4.3           & 0.5     & 16.0         & 6         & 0.08   \\
Y1$\zeta $- & Y  & 5           & 30.2        & 1.5             & 0.1             & Yes  & Adap   & 5               & 1.5            & 4.3           & 0.5     & 8.0          & 6         & 0.04   \\
Y1f$\sigma$ & Y  & 5           & 30.2        & 1.5             & 0.1             & Yes  & Fix    & 5               & 1.5            & 4.3           & 0.5     & 8.0          & 10        & 0.08   \\
Y2          & Y  & 5           & 30.2        & 1.5             & 0.1             & Yes  & Adap   & 5               & 2.5            & 2.5           & 0.0     & 8.0          & 6         & 0.08   \\ 
Y2Mb-       & Y  & 5           & 30.2        & 1.5             & 0.1             & Yes  & Adap   & 3               & 2.5            & 2.5           & 0.0     & 8.0          & 6         & 0.08   \\ 
Y2Mb+       & Y  & 5           & 30.2        & 1.5             & 0.1             & Yes  & Adap   & 7.5             & 2.5            & 2.5           & 0.0     & 8.0          & 6         & 0.08   \\ 
Y4f$\zeta$- & Y  & 5           & 30.2        & 1.5             & 0.1             & Yes  & Fix    & 5               & 1.5            & 2.2           & 0.5     & 8.0          & 6         & 0.04   \\
YG1	    & YG & 5           & 30.2        & 1.5             & 0.1             & Yes  & Adap   & 5               & 1.5            & 4.3           & 0.5     & 8.0          & 6         & 0.08   \\ 
YN1         & Y  & 5           & 30.2        & 1.5             & 0.1             & No   & Adap   & 5               & 1.5            & 4.3           & 0.5     & 8.0          & 6         & --     \\
A2$\tau$    & A  & 10          & 30.2        & 1.5             & 0.8             & Yes  & Adap   & 5               & 2.5            & 2.5           & 0.0     & 8.0&$6 + 30\e^{-t/1.5\gyr}$ & 0.08   \\
E2          & E  & 15          & 30.2        & 2.5             & 1.2             & Yes  & Adap   & 5               & 2.5            & 2.5           & 0.0     & 8.0          &  6        & 0.08   \\
F2          & F  & 5           & 51.7        & 1.5             & 0.1             & Yes  & Adap   & 5               & 2.5            & 2.5           & 0.0     & 8.0          &  6        & 0.08 \\ \hline

  \end{tabular}
  \label{modeltable}
\end{table*}

\section{Simulations}
\label{sec:simulations}

The simulations analysed in this paper are a subset of the models presented
in ABS16. These are simulations of growing disc galaxies within non-growing
live dark matter haloes made using the Tree Smoothed Particle Hydrodynamics
(TreeSPH) code GADGET-3, last described in \citet{gadget2}.  We focus on
standard-resolution models, which contain $N=5\times10^6$ particles in the 
final disc and the same number of halo particles.  Most of the simulations are
collisionless, but a subset contains an isothermal gas component with
pressure $P=\rho c_s^2$ and sound speed $c_s=10\kms$. The global gas fraction in
these discs is kept roughly constant over time at $f_g=10$ per cent. In addition,
most simulations contain a population of short-lived, massive
particles representing GMCs. The force softening lengths are $\epsilon_{\rm
bar}=30\pc$ for baryonic particles (including GMCs) and $\epsilon_{\rm
DM}=134\pc$ for DM particles.

Table~\ref{modeltable} gives an overview of the models cited here.
We give only a brief account of the meaning of the model parameters -- a full
description can be found in ABS16. Standard models contain GMCs, but no gas 
and all models have evolved over a simulation time of $t_{\rm f}=10\gyr$. 
The presence of gas in a model is marked by a `G' in its name, while the 
absence of GMCs is marked by an `N'.  

The initial conditions (ICs) were created using the GALIC code \citep{yurin}. 
The details of the ICs can be found in Table 1 of ABS16.
The models discussed here all start with a spherical 
dark matter halo with a \citet{hernquist} profile and a mass of $10^{12}\msun$. 
The F model differs from the others in that the scale length of its halo is 
$a_{\rm halo}=51.7\kpc$ rather than $30.2\kpc$. 

All models analysed here contain an IC disc with a mass profile
\begin{equation}
{\rho_{\rm{disc,i}}(R,z)} = {{M_{\rm{b,i}}}\over{4\pi {z_{0,{\rm
disc}}}{h_{R,{\rm disc}}}^2}} {\sech^2 \left({z}\over{z_{0,{\rm
disc}}}\right)} {\exp\left(-{R}\over{h_{R,{\rm disc}}}\right)}.
\end{equation}
Here $h_{R,{\rm disc}}$ is the IC disc exponential scalelength and
a radially constant isothermal vertical profile with scaleheight $z_{0, {\rm disc}}$
is assumed. The Y and F models start with a baryonic disc of mass 
$M_{\rm b,i}=5\times10^9\msun$, which is compact ($h_{R,{\rm disc}}=1.5\kpc$)
and thin ($z_{0, {\rm disc}}=0.1\kpc$). The A models contain a thicker and more 
massive IC disc ($z_{0, {\rm disc}}=0.8\kpc$, $M_{\rm b,i}=10\times10^9\msun$) 
and the IC disc in the E models is even more massive, thicker and more extended
($h_{R,{\rm disc}}=2.5\kpc$, $z_{0, {\rm disc}}=1.2\kpc$, $M_{\rm b,i}=15\times10^9\msun$).

Stellar particles are continuously added to the disc on near-circular orbits.
The young stellar populations are assigned birth velocity dispersions
$\sigma_0$ in all three directions $R$, $\phi$ and $z$. The standard choice is
$\sigma_0=6\kms$, but we also consider a simulation with $\sigma_0=10\kms$
(Y1f$\sigma$) and one (A2$\tau$) in which the birth velocity dispersion declines 
exponentially with time 
\begin{equation}\label{sigmazerot}
\sigma_0(t)=\left(6+30\e^{-t/1.5\!\gyr}\right)\kms.
\end{equation}
The star-formation rate is
\begin{equation}\label{eq:SFTt}
{\rm SFR}(t)={\rm SFR}_0 \times \exp({-t/t_{\rm SFR}}),
\end{equation}
with $t_{\rm SFR}=8$ or $16\gyr$. The constant ${\rm SFR}_0$ is adjusted to produce at $t=t_f$
a target final baryonic mass $M_{\rm f}$ in the range $3-7.5\times10^{10}\msun$.
Mass growth is smooth in time and sufficiently slow for the process to
be effectively adiabatic.

Particles are added randomly distributed in azimuth every five Myr 
with an exponential radial density profile $\Sigma_{\rm SF}(R)\propto\exp(-R/h_R(t))$.
The scalelength $h_R(t)$ of the newly added particles grows in time as
\begin{equation}\label{eq:hRt}
h_R(t)=h_{R,\rm i}+(h_{R,\rm f}-h_{R,\rm i})(t/t_{\rm f})^\xi.
\end{equation}
To avoid inserting particles in the bar region, where near-circular orbits
do not exist, particles are not added inside the cutoff radius $R_{\rm cut}$,
which is either determined by the current bar length (`adaptive cutoff'), or given by a
pre-determined formula $R_{\rm cut}(t)=\left(0.67+{0.33t\over1\,{\rm Gyr}}\right) \kpc$
(`fixed cutoff').

GMCs are modelled as a population of massive particles drawn from a mass
function of the form ${\rm d}N/{\rm d}M\propto M^\gamma$ with lower and upper mass 
limits $M_{\rm low}=10^5\msun$ and $M_{\rm up}=10^7\msun$ and a power law exponent
$\gamma=-1.6$. Their radial density is proportional to $\Sigma_{\rm SF}$, 
and their azimuthal density is given by
\begin{equation}
\rho_{\rm GMC}(\phi)\propto \left[\rho_{\rm ys}(\phi)\right]^\alpha,
\end{equation}
where $\rho_{\rm ys}$ is the density of young stars and $\alpha=1$. The
mass in GMCs is determined by the SFR efficiency $\zeta$.  Specifically, for
each $\Delta m_{\rm stars}$ of stars formed, a total GMC mass $\Delta
m_{\rm GMC} = \Delta m_{\rm stars}/\zeta$ is created. GMC particles live for
only $50\myr$: for $25\myr$ their masses grow with time, and for the
final $25\myr$ of their lives their masses are constant.

In ABS16 we presented an overview of the properties of the simulated
galaxies. Important findings of ABS16 that are relevant to the results
of this paper, are:
\begin{enumerate}
\item{In the absence of GMCs, the models are too cold vertically to
explain the vertical profile of the MW thin disc. Heating by GMCs creates
remarkably exponential vertical profiles, the scaleheights of which agree roughly with that
inferred for the MW thin disc for our standard GMC mass function and $\zeta=0.04$ (Y1$\zeta$-).
These discs have radially constant vertical profiles.}
\item{GMC heating is particularly efficient early on, when SFRs are high and stellar 
disc masses are small.}
\item{No thick discs similar to the one observed in the MW form in the models with thin IC discs. Thicker
discs can form in models with high baryon fractions (Y4f$\zeta$-, F2), but they are too hot radially.}
\item{Spurious two-body heating due to halo particles is negligible when the halo
is resolved with at least $5\times10^{6}$ particles.}
\item{The output scalelength $h_R$ of a stellar population can differ significantly from the 
input scalelength, as bars and spirals drive radial redistribution and lead to an increase
of $h_R$ in the outer disc.}
\item{Bars are stronger for more compact models and weaker for models with an isothermal gas component.}
\item{Isothermal gas components mildly increase the efficiency of GMC heating, presumably
as GMC particles attract wakes of gas which increase their effective mass.}
\end{enumerate}

Unless otherwise noted, we will analyse our models at a solar-like radius $R_0=8\kpc$. The local 
exponential scalelength $h_R$ at $R_0$ can differ significantly between models (2-5 kpc), but the 
extreme values are caused by deviations from simple exponential profiles. The differences in 
final surface density at $R_0$ are small (Fig. 12 in ABS16) and for our standard
disc mass the models agree reasonably with estimates for the Snhd \citep{flynn}.

\section{The impact on AVRs of biases and errors in ages}
\label{sec:biases}

\begin{figure*}
\centering
\vspace{-0.cm}
\includegraphics[width=18cm]{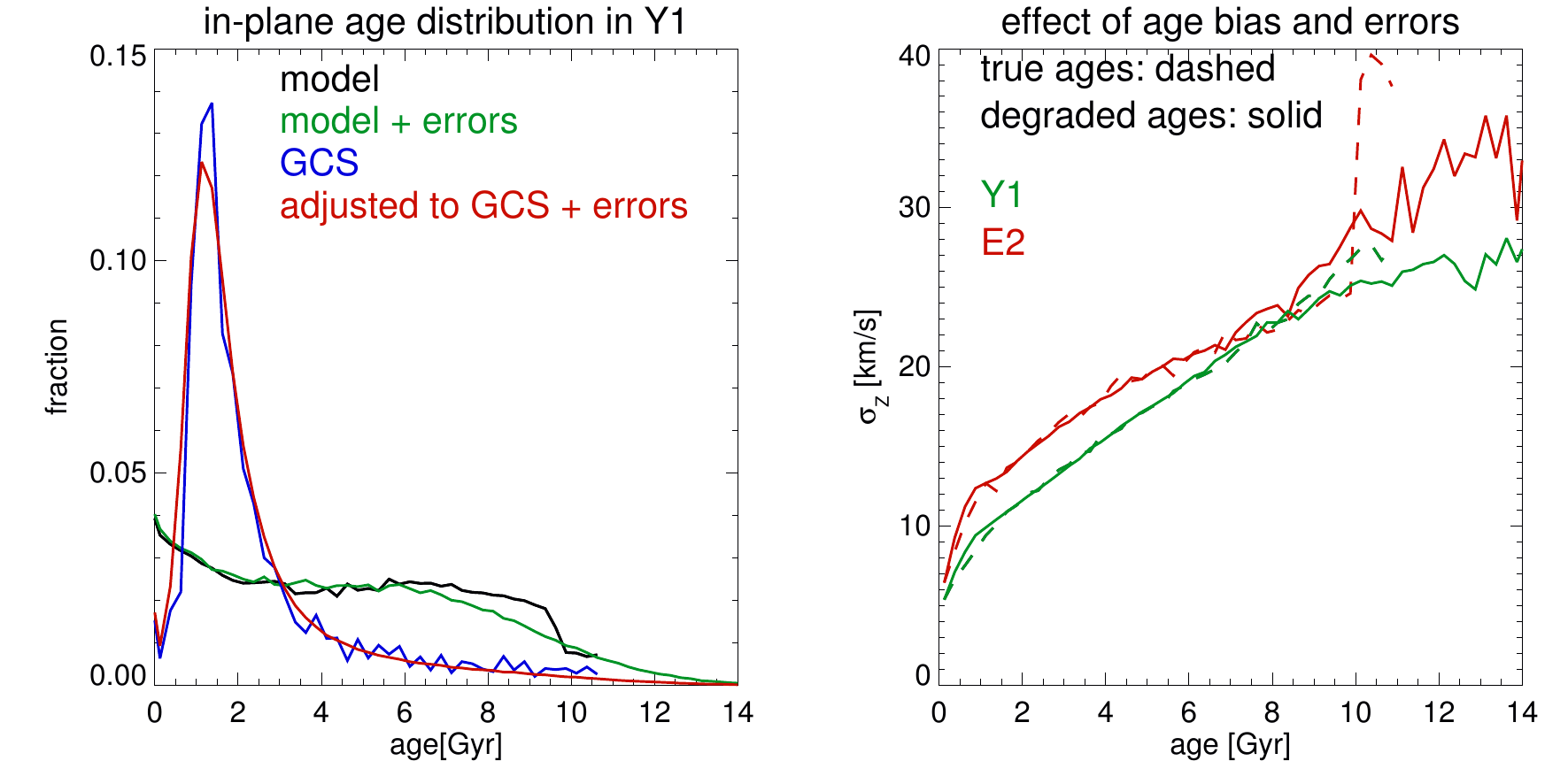}\\
\caption
{Correcting for age bias and errors to compare models to the GCS data. Left: 
The black curve shows the age distribution of stars in Model Y1 at 10 Gyr and
at $R=8\pm0.5\kpc$ and $z=0\pm0.1\kpc$. If we assume observational $1\sigma$ 
errors of 20 per cent we predict the green line. The blue line shows the age
distribution in the GCS sample. The red line is what we get when we weight stars
in the model so that they match the blue line and then assume 20 per cent age
errors. Right: Vertical AVRs $\sigma_z(\tau)$ for 
Models Y1 (green) and E2 (red) at $10\gyr$ and at $R=8\pm0.5\kpc$ and $|z|\le0.1\kpc$.
Dashed lines are for the true age distribution (black line in left panel) and 
solid lines are for the degraded age distribution (red line in left panel).}
\label{gcs}
\end{figure*}

Before we can compare the AVRs from the \emph{Geneva-Copenhagen Survey} (GCS)  
\citep{nordstroem, casagrande} to
data from simulations, we need to model the impact on the observations of the
survey's selection function and errors in the estimated ages of stars
(see also \citealp{holmberg7, martig}).  

We start by selecting GCS stars with $\feh>-0.8$ and heliocentric azimuthal
velocity $V>-150\kms$ and a `good' age determination: from Casagrande et al., we use the
maximum likelihood ages $\tau$ and the ages $\tau_{16}$ and $\tau_{84}$ of the 
$16$ and $84 \%$ quantiles of the probability distribution in age.  If either
$\tau_{84}-\tau_{16} < 2\gyr$ or $2\left(\tau_{84}-\tau_{16}\right) /
\left(\tau_{84}+\tau_{16}\right) < 0.5$ the age $\tau$ is deemed good and the
star enters our sample. These $\sim7500$ stars are ordered by age and placed
in bins with 200 stars each to calculate $\sigma(\tau)$. Every 10 stars a new point
is plotted and thus every 20th point is statistically independent of its predecessors.

The blue curve in
Fig.~\ref{gcs} shows the strongly biased age distribution of this sample. 
In addition to the bias of this mid-plane survey towards kinematically colder
and hence younger stars, the GCS is largely magnitude limited, favouring 
more luminous, younger stars. The exclusion of hot/blue stars with 
$T_{\rm eff} \gtrsim 7000 {\rm K}$ biases the sample against very young stars, 
and in particular excludes massive stars that could have safe age-determinations
with $\tau \lesssim 1.5 \gyr$. The small upturn in the number densities at 
$\tau < 0.5 \gyr$ can be mostly ascribed to a pile-up of maximum likelihood 
values from main-sequence stars with temperature and/or metallicity over-estimates,
and hence underestimated maximum likelihood ages. Consistently, there is little
evolution in GCS velocity dispersions for ages below $\tau \sim2\gyr$, stalling
at values typical for $\sim 1.5 \gyr$ old stars and far above the values 
derived by AB09 for their bluest stars.

Blue \emph{Hipparcos} stars can be used to determine velocity dispersions of
very young stars. The lowest dispersions found by AB09 for their bluest bins
in $B-V$ are $\sigma_z=5.5 \kms$ and $\sigma_R=8 \kms$. These stars will,
however, be kinematically biased, as a majority of them belong to a small number of
moving groups of young stars. It is interesting that $\sigma_R$ increases
very strongly between $B-V=-0.2$ and $B-V=0$, so the smallest value of
$\sigma_z/\sigma_R\sim1/3$ occurs at $B-V\sim0$.

We impose the GCS age bias and errors of $\sim 20$ per cent in ages on data from
a model galaxy as follows: first we select star particles at
$R=8\pm0.5\kpc$ and $|z|\le 0.1\kpc$, which approximates the GCS volume.
Stars already present in the ICs are randomly assigned ages $\tau \in \left[t_{\rm
f}, t_{\rm f} + 1 \gyr \right]$. To each star $i$ we assign a weight $w_i\ge1$
so that the distribution of true ages of the weighted sample agrees with the 
one of the GCS sample (blue curve in Fig.~\ref{gcs}). We then determine $w_i$ 
`observed' ages for each star by assuming a Gaussian error distribution with 
standard deviation 20 per cent of the true age. The resulting distribution
of all assigned ages is shown as the red curve in Fig.~\ref{gcs}. For the 
models all bins have width $\Delta\tau=0.25\gyr$. Note that the GCS selection function
also depends on metallicity, which is not taken into account here.
 
The black curve in the left panel of Fig.~\ref{gcs} shows for Model Y1 at
$t_{\rm f}=10\gyr$ the actual age distribution of `solar neighbourhood' stars. This
distribution peaks at the youngest ages because young stars are confined to
the plane, where we select particles.  At intermediate ages the distribution
is rather flat because heating and inside-out growth are balanced by the
declining SFR.  The oldest stars are underrepresented as they are hot and
centrally concentrated. The green curve is obtained on folding the black
curve with our assumed errors in age.  Now stars with ages above
$5\gyr$ are smeared into a tail that extends to $14\gyr$. The striking
difference between this green curve and the blue curve for the age
distribution of our GCS sample shows how strongly the survey's selection
function biases the data.

In the right panel of Fig.~\ref{gcs} we show the derived AVRs $\sigma_z (\tau)$ for Models 
Y1 (green) and E2 (red).  The dashed lines show 
the AVRs for true age distributions, and the
solid lines the AVRs for the GCS-like samples just described.  The
jump at $10\gyr$ in the red dashed curve reflects the hot proto-thick disc
included in the ICs of Model E2. The main difference between the dashed and
full curves is extension of the end point from 10 to $14\gyr$ and elimination
of the step in the dashed curve for E2 (see also \citealp{martig}). At low ages 
the solid curves lie slightly above the dashed curves as a consequence of stars 
with ages $\sim2\gyr$ being scattered to lower `observed' ages.

\section{AVRs and velocity ellipsoid shapes}
\label{sec:AVR}

\begin{figure*}
\vspace{-0.cm}
\includegraphics[width=18cm]{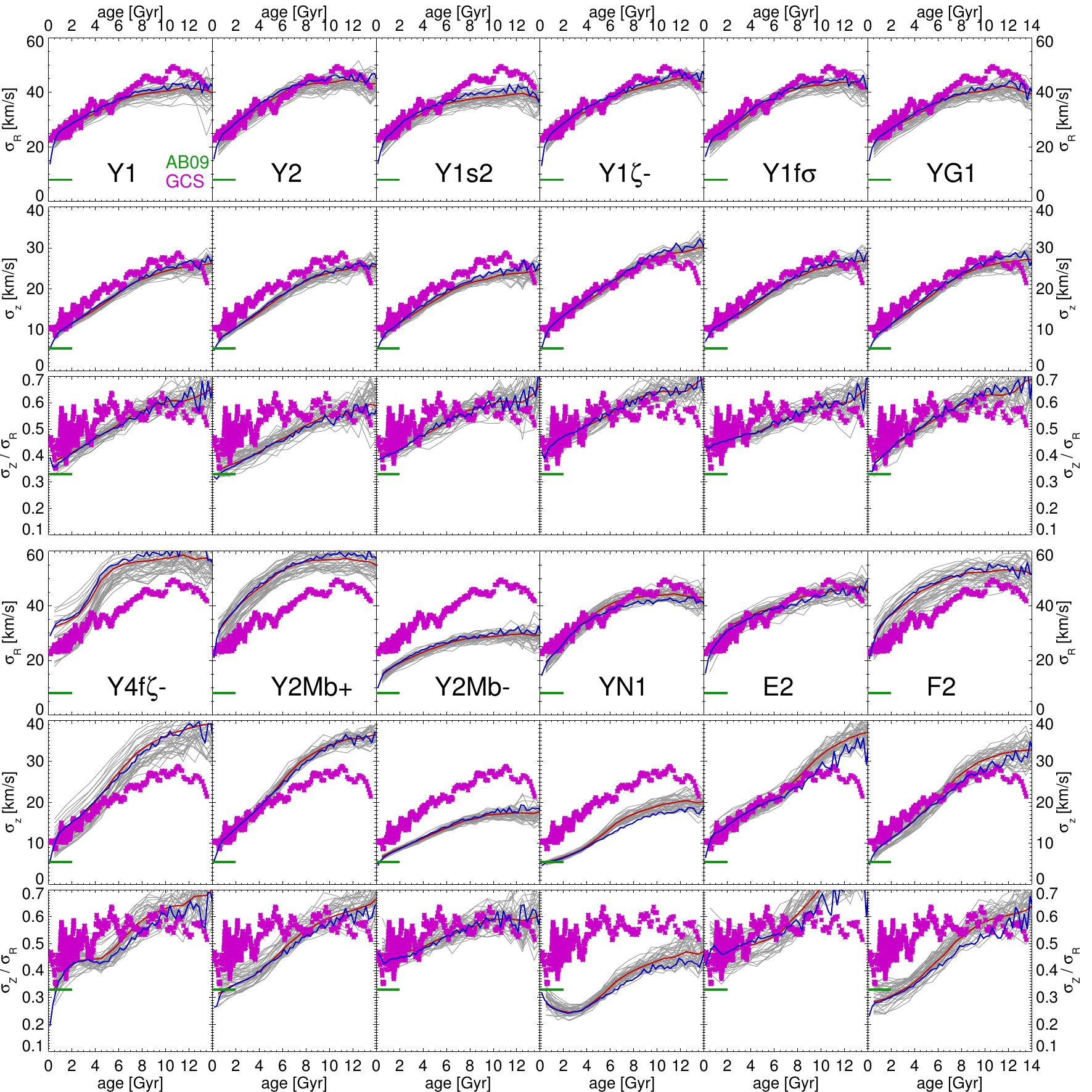}\\
\caption
{The AVRs of a variety of models. For each models 
we show $\sigma_R(\tau)$ (top), $\sigma_z(\tau)$ (middle) and 
$\sigma_z/\sigma_R (\tau)$ (bottom). The pink asterisks are extracted from the 
GCS Snhd data of \citet{casagrande}. The blue lines are the azimuthally averaged
AVRs from the models for $|z|<0.1\kpc$ and the red lines are for $|z|<0.3\kpc$. 
The set of grey lines are for 36 equally spaced azimuthal bins at $|z|<0.3\kpc$.
All model data are for $R=8\pm0.5\kpc$. The green lines at low ages mark the 
lowest values found by AB09 for each quantity with blue \emph{Hipparcos} stars.
The GCS stars with good ages are placed in bins with 200 stars each and every 10 stars a new point
is plotted, so that every 20th point is statistically independent.
At $\tau\sim 2\gyr$, there are 200 stars per $\sim100\myr$, whereas at $\tau>6\gyr$
there are only $\sim 1000$ stars in total.}
\label{azi}
\end{figure*}

In the last section we explained how we extract comparable AVRs from the
models and the GCS. Here we compare $\sigma_z(\tau)$,  $\sigma_R(\tau)$ 
and the ratio $\sigma_z/\sigma_R (\tau)$ between models and the GCS sample.

\subsection{Azimuthal variation of the AVR}
\label{sec:azimuth}

Whereas in our models we can select star particles from every azimuth, all 
GCS stars are drawn from azimuths near that of the Sun. Just as the density 
of stars in disc galaxies varies with azimuth, we expect 
the age and velocity distributions of stars to vary as well. Moreover, a 
bar or spiral arm drives non-axisymmetric streaming (e.g. \citealp{dehnen}),
and streaming velocities will boost the recovered velocity dispersion when
star particles are binned regardless of azimuth. 

Here we attempt to quantify the azimuthal variation of the AVR in our models.
We make no attempt to take into account the actual inter-arm location of the Sun
because this task requires (i) reliable knowledge of the Galaxy's
non-axisymmetries, (ii) precise control of the non-axisymmetries within the
models, (iii) sufficient star particles within $\sim0.1\kpc$ of the Sun to make
Poisson noise unimportant, and (iv) detailed modelling of the selection function.
None of these three conditions being satisfied, we
confine ourselves to the estimation of the uncertainties arising from
uncharted non-axisymmetric structures.

For each model we divided the annulus $R=7.5-8.5\kpc$ at
$\left|z\right|<0.3\kpc$ into 36 sectors of 10 deg width and determined
mean motions and dispersions for each sector. These 36 separate AVRs are
displayed as grey lines in Fig.~\ref{azi}.  The area occupied by these
lines should be regarded as the region within which the AVR of a mock Snhd
could fall.  On average we find that higher velocity dispersions are
found in regions of higher star density, but the scatter is significant.  For
most models a typical azimuthal spread in $\sigma_z(\tau)$ is $\sim \pm 1
\kms$ at young ages and $\sim2-3\kms$ at older ages.
Models such as Y4f$\zeta$- and E2 with bars that nearly reach $R=8\kpc$, show
larger spreads, even at young ages, because bars have strong effects on
stellar orbits and significant differences in orbit populations occur in
regions positioned differently with respect to the bar.  Bars and spiral
structure excite streaming motions more parallel to the plane than
vertically, so the fractional azimuthal spreads are larger for
$\sigma_R(\tau)$ than for $\sigma_z(\tau)$.

In Fig.~\ref{azi} we also plot in red the azimuthal average for
$|z|<0.3\kpc$ and in blue for $|z|<0.1\kpc$, which is the most relevant region
to compare to the GCS data. Unfortunately, when we subdivide the data by
azimuth, Poisson noise is unacceptably large in the data for $|z|<0.1\kpc$.
The biggest discrepancies between red and blue lines are found for
$\sigma_z(\tau)$ in Models E2 and F2, which have thicker discs, and
in Model YN1, which has no GMCs. The discrepancy indicates vertical
dispersions which increase with altitude at a given age, as the red lines probe
higher altitudes.  For most models dispersions hardly change with altitude
and red and blue lines are very similar indicating that our use of bins
extending to $|z|=0.3\kpc$ will  not mislead.

In Fig.~\ref{azi} the Snhd data from the GCS are shown in pink.  Green horizontal lines
at low ages indicate the lowest values found for each of the three
quantities for blue \emph{Hipparcos} stars by AB09 to give an indication of
the uncertainties at low ages. 

\subsection{Specific models}\label{sec:specifics}

We begin by analysing models, which ABS16 considered to have problematic AVRs
to reconsider this judgement in light of the azimuthal spreads shown in
Fig.~\ref{azi}. 

Models Y2Mb+ and Y2Mb- have an abnormally high and low final disc mass,
respectively.  In these models the azimuthal variations are too small to
account for the conflict with observation.  Similarly, the vertical
dispersions in Model YN1, which lacks GMCs, remain too small and the
dispersions in the thickest Y model, Y4f$\zeta$-, remain too high.  Finally,
the characteristic feature in the radial dispersion at $\tau\sim 4 \gyr$ for
Y4f$\zeta$- shows up clearly at all azimuthal positions.

ABS16 showed that models that start from IC F, which has a low-density dark
halo, produce thicker discs, which are, however, too hot.  In particular
Model F2 has a vertical profile that is significantly thicker than those of
our standard models, run from initial condition Y, and is slightly
double-exponential, but still significantly thinner than that of the MW.
When the azimuthal variation in its AVRs is taken into account, Model F2
becomes marginally acceptable because at the right end  of the lower
row of panels in Fig.~\ref{azi}, the lowest grey curve for
$\sigma_R(\tau)$ is consistent with the observations.

We now consider models run from the E and Y ICs. These models have standard dark halos
and total baryonic masses $M_{\rm f}=5\times10^{10}\msun$.  Model Y1 is
slightly too cold radially for old stars, and slightly too cold vertically
for stars of intermediate age. The more compact Model Y2 has a radial AVR
that fits observations better because it has higher surface densities and
thus stronger non-axisymmetric structures. In Model Y1s2 the SF timescale is
longer so the disc grows more slowly. The consequence is old populations
that are too cold radially and vertically because the total mass and the GMC
mass fraction at early times are both low. Model Y1f$\sigma$, which uses a
fixed cutoff and has a higher value for the input velocity dispersion
$\sigma_0$, shows a mildly better fit to the data than Y1, as does Model YG1,
which has an isothermal gas component. The best fits by a Y model are
provided by Model Y1$\zeta$-, which has a lower SF efficiency and thus a
higher total GMC mass at all times.

Whereas the Y models start from a small, compact and thin disc, Model
E2 starts from a more extended and more massive thick disc.  If we consider
that we should not trust any points at $\tau>12 \gyr$ and that the blue lines
in Fig.~\ref{azi} are more relevant than the red lines, E2 also provides
acceptable fits to both $\sigma_R(\tau)$ and $\sigma_z(\tau)$. At old ages
its AVR $\sigma_z(\tau)$ differs from that of Model Y2 on account of the
thick IC, and at young ages it differs on account of its extended and
buckling bar.

\subsection{Shape of the velocity ellipsoid}
\label{sec:arat}

We now consider $\sigma_z/\sigma_R (\tau)$, which is plotted for each model
in the third and sixth rows of Fig.~\ref{azi}.  The (pink) observational
values show substantial scatter around $\sigma_z/\sigma_R\sim 0.5$, with
lower values at young ages and higher values at old ages. Although the
\citet{casagrande} values lie in $(0.4,0.6)$, AB09 found values as low as
0.33 for the bluest stars in the \emph{Hipparcos} catalogue, which, as we
noted above, are excluded from the GCS sample, but may be biased by moving 
groups. At old ages, $\tau \gtrsim 7 \gyr$, $\sigma_z/\sigma_R\approx0.55-0.6$
appears rather constant. Values of this order are predicted by simulations 
of heating by GMCs \citep{haenninen, sellwood}.

In all models stars are, by construction, added with
$\sigma_z=\sigma_R=\sigma_0$. Non-axisymmetries almost instantaneously increase in
$\sigma_R$ to a value $\gg\sigma_0$, while $\sigma_z$ increases much more
gradually. Consequently, quite soon $\sigma_z/\sigma_R<0.5$, as observed for
the youngest stars.  Moreover, $\sigma_z/\sigma_R$ increases with age, again
as observed.

Although Model Y2Mb- with a low-mass disc provides a very good fit to the observed
$\sigma_z/\sigma_R (\tau)$, its velocity dispersions are too low to be
consistent with observations. Model F2 with a low-density halo and a
marginally acceptable AVR,  can now be excluded because in it
$\sigma_z/\sigma_R (\tau)$ is too low at young ages.

The E and Y models have standard dark haloes. In Model Y1 $\sigma_z/\sigma_R
(\tau)$ is lower than in the observations at $\tau\sim5\gyr$,
and higher for $\tau\ga10\gyr$.  Model Y2 with a more compact disc fits
$\sigma_z/\sigma_R$ less well. Model Y1f$\sigma$, which has an abnormally
high value of the birth dispersion parameter $\sigma_0=10\kms$, fits the observations
better. Model Y1$\zeta$-, which has an abnormally low star-formation
efficiency, provides an excellent fit except at ages $\tau\ga8\gyr$.  Model
Y1s2, which has a more slowly declining SFR, provides a similar quality of
fit to that of Model Y1.

Model E2, which has a massive and extended primordial thick disc, provides a
good fit to the observed $\sigma_z/\sigma_R(\tau)$ at $\tau<9\gyr$, but
provides unacceptably high values at older ages. However, this conclusion must
be moderated by two caveats: (i) the grey lines should be moved downwards by
the separation between the blue line for $|z|<0.1\kpc$ and the red line for
$|z|<0.3\kpc$; (ii) no model has stars with $\tau>11\gyr$ (ICs stars are assigned
ages at 10-11 Gyr), so we must be careful when drawing conclusions regarding
this age range. In light of these caveats, we consider that Model E2 also
provides an acceptable fit to $\sigma_z/\sigma_R (\tau)$.

As expected for a model without GMCs, $\sigma_z/\sigma_R(\tau)$ is too low at all
ages $\tau$ in YN1. We note that the increase in $\sigma_z/\sigma_R(\tau)$ for older
stars may in part be caused by spurious collisional relaxation \citep{sellwood13}.

\begin{figure*}
\centering
\includegraphics[width=18cm]{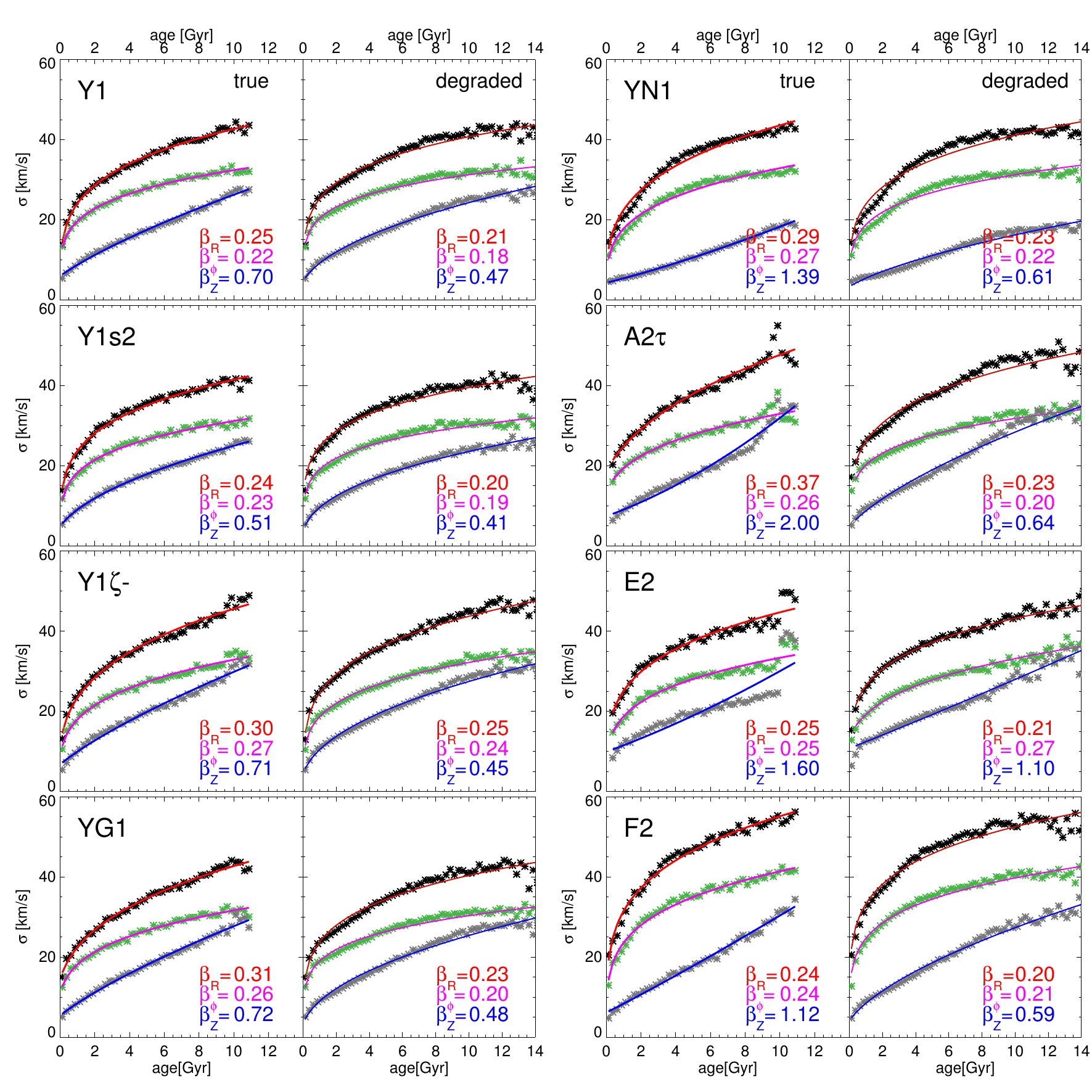}\\
\caption
{Asterisks display AVRs for all three directions $R$ (black), $\phi$ (green) 
and $z$ (grey) at $R=8\pm0.5\kpc$, $z=0\pm0.1\kpc$ and  $t=10\gyr$ for several models. Overplotted are fits of the form 
$\sigma_{10}\left[\left(\tau+\tau_1\right)/ \left(10\gyr + \tau_1\right)\right]^{\beta}$
in red ($R$), pink ($\phi$) and blue ($z$). For each model we show AVRs for both 
true (left) and degraded ages (right) according to Section \ref{sec:biases}. The
fitted values for $\beta$ are displayed in corresponding colours in the bottom 
right of each panel.
}
\label{heatfits}
\end{figure*}

\subsection{Which models can reproduce the Snhd AVR?}
\label{sec:whichModel}

Combining the results of all three quantities plotted in Fig.~\ref{azi}, we
conclude that only models with a standard halo (Y and E models) and standard
baryonic mass ($M_{\rm f}=5\times10^{10}\msun$) are compatible with the Snhd
data. Lower or higher disc masses, or lower density haloes, all fail to
reproduce the data. As far as models with thicker disc components are
concerned, inclusion of a thick disc in the ICs (Model E2) is clearly
favoured over a thick disc that emerges during the simulations (Models F2 and
Y4f$\zeta$-).

Since our models include significant idealizations, we cannot
expect any model to provide a perfect fit to the data and we do not seek to
reproduce the MW disc precisely.  Not only do the models still lack heating
processes, such as interactions with satellite galaxies (e.g.
\citealp{just}), that will modify the predictions, but even after our effort
to take age biases and errors into account, the comparison of data from
models and observations must be imperfect.  Moreover, the resolution of our
models is still too low and the history of the MW is too rich in events to
allow the detailed modelling of velocity distributions of stars in the $\sim
100\pc$ sphere covered by the GCS.

Considering all the shortcomings listed above and despite their failure to
model properly the thick disc component of the MW, models such as Y1$\zeta$-
or E2 provide very good fits to the data. We thus emphasise the conclusions
already drawn in ABS16: (i) combined disc heating by GMCs and
non-axisymmetries in the disc is very likely responsible for the overall
shape of the AVR in the Snhd; (ii) the models clearly favour a 
baryonic disc mass $5\times10^{10}\msun$ and the cosmologically inferred dark-halo
mass parameters.

\subsection{Power-law indices of AVRs}
\label{sec:heatIndex}

We now examine the values of $\beta$ that are obtained by fitting AVRs 
to equation \eqref{eq:heatlaw} with $x=\tau$. Small values indicate larger
differences in the velocity dispersions of young stars and stars of
intermediate ages than between the stars of intermediate and old ages.
Over the last half century
values of  $\beta$ between 0.25 and 0.6 have been found
from various samples of the Snhd stars (see e.g. Table 1 in
\citealp{haenninen}). Data from the \emph{Hipparcos} and
\emph{GCS} surveys indicate that $\beta_z\sim0.45-0.53$ is
larger than $\beta_R\sim0.31-0.39$ (\citealp{holmberg}, AB09).  In view of
the influence of age biases and errors on empirical AVRs, the spread of
values is not surprising.

Fig.~\ref{heatfits} displays for the endpoints of several models the AVRs for the
stars in the annulus $R=8\pm0.5\kpc$ and at $|z|<0.1\kpc$. There are two panels
for each model because the AVRs using true ages are plotted in the left
panel, while the right panel shows the AVRs yielded by ages degraded as
described in Section~\ref{sec:biases}.  The model data are displayed as black
($\sigma_R$), green ($\sigma_\phi$) and grey points ($\sigma_z$).  Fits of
equation \eqref{eq:heatlaw} to these data are over-plotted in red
($\sigma_R$), pink ($\sigma_\phi$) and blue ($\sigma_z$).  The corresponding
heating exponents $\beta_i$ are displayed in the lower right corner of each
panel. 

In this section we analyse also the azimuthal velocity dispersion
$\sigma_{\phi}$. For the most part we exclude $\sigma_{\phi}$ from our
analysis because (i) the skewness of the $v_{\phi}$ distribution (see e.g.
\citealp{gd2}, Section 4.4.3) renders $\sigma_\phi$ hard to interpret, and
(ii) it is tightly coupled by dynamics to $\sigma_R$, so it does not provide
independent diagnostic information. Models yield $\sigma_{\phi}/\sigma_{R}\sim
0.75-0.8$ at $R=8\kpc$ and $|z|<0.1\kpc$, whereas observations of the Snhd
yield  smaller values,  $\sigma_{\phi}/\sigma_{R}\sim 0.65$.  Azimuthal
variation can lower the observed value by $0.05-0.1$ for some azimuths, which
only brings some models to marginal agreement with the Snhd data.  We
interpret this discrepancy as a result of selection biases. Any selection
bias exerting a preference in metallicity has a tendency to lower
$\sigma_{\phi}/\sigma_{R}$. This kind of bias has been discussed in
\citet{sbd10}.  We leave the detailed investigation of this phenomenon to a
future paper.

Fig.~\ref{heatfits} shows that the data can usually be nicely fitted by equation
\eqref{eq:heatlaw}. Notable exceptions are, as expected, the AVRs from true
ages for Models A2$\tau$, which features a declining input dispersion
(eq.~\ref{sigmazerot}), and E2 which has a thick
disc in its IC. Consequently, in both models all three dispersions, but especially
$\sigma_z$, have sharp upturns at the oldest ages.

The true AVRs of the other models have heating exponents in the range
$\beta_R=0.24-0.31$, $\beta_{\phi}=0.22-0.27$ and $\beta_z=0.51-1.39$. The
in-plane coefficients show hardly any scatter as heating is dominated by
non-axisymmetries, which are similar in all models. $\beta_{\phi}<\beta_R$
holds in all models. AB09 favoured $\beta_{\phi}>\beta_R$ for the Snhd
data but could not exclude $\beta_{\phi}<\beta_R$.

\begin{figure*}
\centering
\includegraphics[width=18cm]{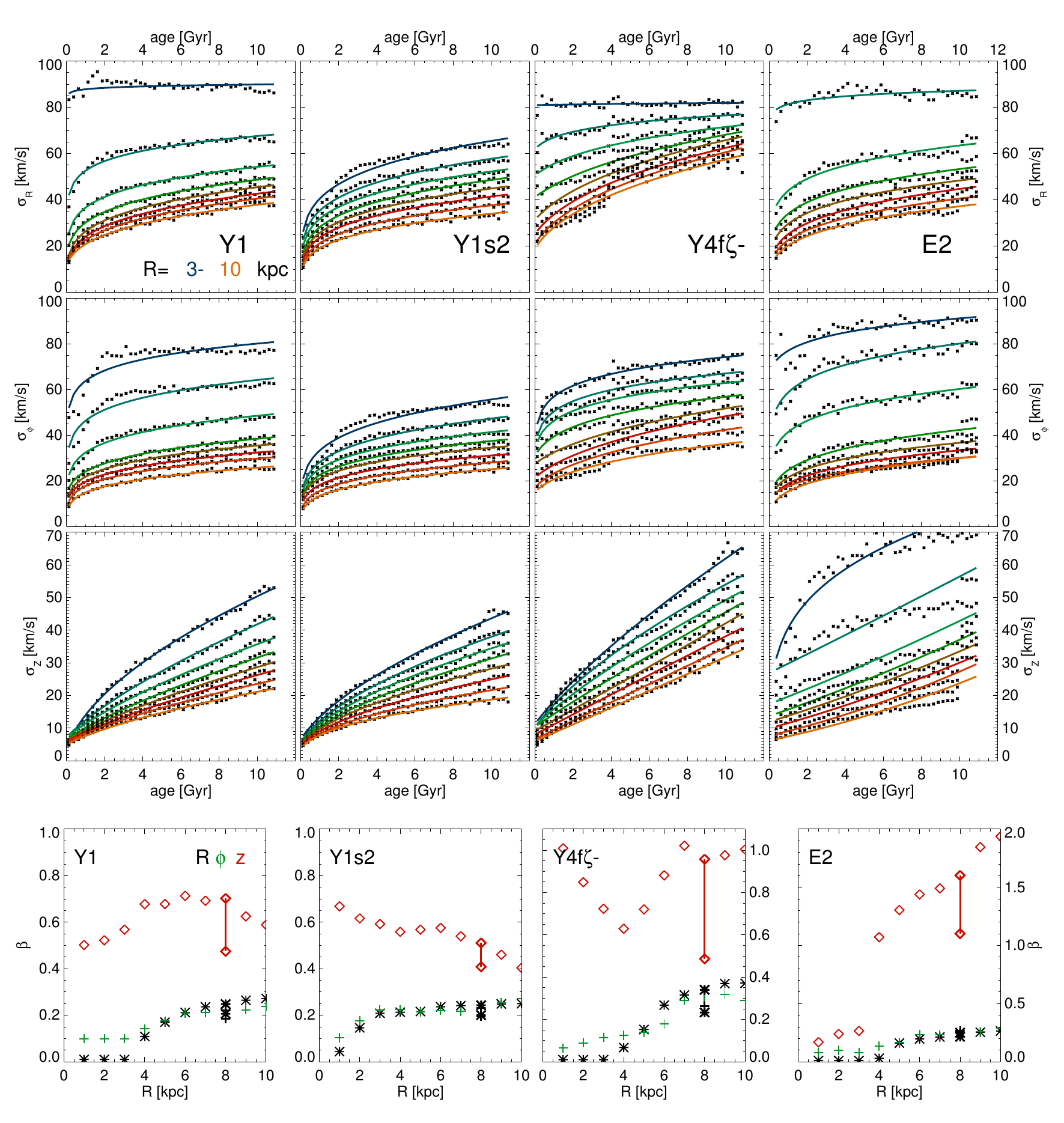}\\
\caption {The AVRs $\sigma_i(\tau)$ at $t=10\gyr$ for true ages and at radii
$R=3,4,\ldots,10\kpc$ and $z=0\pm0.1\kpc$ in directions $R$ (top row), $\phi$ (second row) and
$z$ (third row). Overplotted are fits of the form
$\sigma_{10}\left[\left(\tau+\tau_1\right)/ \left(10\gyr
+\tau_1\right)\right]^{\beta}$.  The bottom row plots the heating parameters
$\beta$ obtained from fits in all three directions as a function of $R$.
Vertical lines at $R=8\kpc$ show how $\beta$ changes when ages are degraded
by errors and observational bias.  } \label{heatrad}
\end{figure*}

The scatter in $\beta_z$ is significant. The lowest value is found for Model
Y1s2. In Section \ref{sec:heathistory}, we will show that this arises from
this model's flatter SF history, which implies a slower decline with time in
the total mass of GMCs. The heating exponent of Model F2 with a low-density
dark halo, $\beta_z=1.12$, is unusually high. In this model an extended bar
forms very early on, which leads to a high inner cutoff radius $R_{\rm cut}$
for the insertion of star and GMC particles. As a consequence, the GMC total
mass at $R=8\kpc$ is at early times higher than in other models and the
decline towards late time is thus stronger. Moreover in this model vertical
heating by the extended $m=2$ non-axisymmetries is non-negligible. Model YN1,
which lacks GMCs, has the highest value of $\beta_z$ because it is heated
vertically by large-scale non-axisymmetries rather than GMCs. We find that
this heating mechanism generally produces higher values of $\beta_z$.

Degradation of the ages flattens AVRs because the observations are
dominated by stars with true ages $\tau\sim 2 \gyr$, which are scattered into
all age bins by observational errors. In our models this flattening is
greatest at the oldest ages, but this is to some degree artificial: we
only have stars with true ages up to $11\gyr$ (IC stars are assigned ages
$10-11\gyr$) and ages can be scattered up $14\gyr$,  so we may be overestimating
the total flattening. However, detailed aspects of the selection function,
such as the exclusion of young blue stars, and of how ages were determined,
for example the restriction $\tau\le14\gyr$ on the ages of GCS stars, 
and of the error distribution, which is significantly non-Gaussian, were 
not modelled here, and could significantly change the parameters of
observed AVRs.

Flattening of the AVR leads to a noticeable reduction in $\beta$, most prominently for
$\beta_z$.  For models with the Y IC that have GMCs (left column in
Fig.~\ref{heatfits}), the reduced ranges are $\beta_R=0.20-0.25$,
$\beta_{\phi}=0.18-0.24$ and $\beta_z=0.41-0.48$. The extremely high values
of $\beta_z$ for Models YN1 and F2 are reduced to 0.61 and 0.59, respectively.
For Models A2$\tau$ and E2 that have hot old components in their ICs,
degradation of the ages smears out the AVRs at old ages and leads to better
fits by equation \eqref{eq:heatlaw}.  Nonetheless, even after flattening 
$\beta_z=0.64$ for Model A2$\tau$ is high, and for Model 
E2 $\sigma_z(\tau)$ is fitted best by an almost straight line: $\beta_z=1.1$.

\begin{figure*}
\centering
\includegraphics[width=18cm]{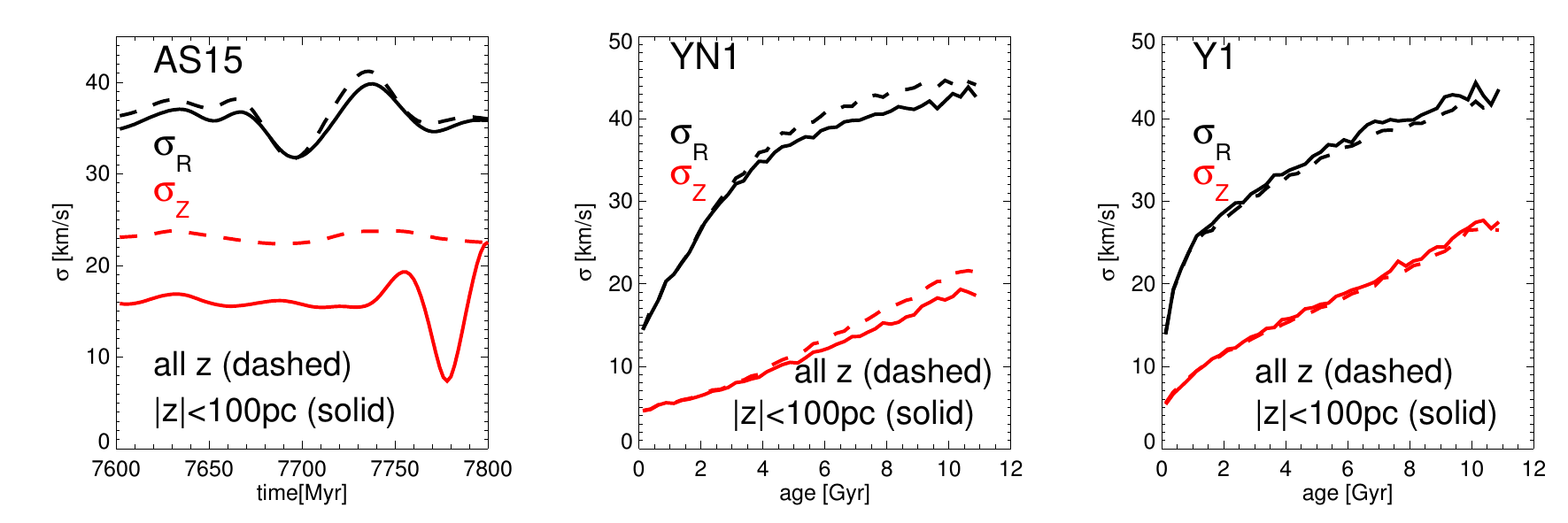}\\
\caption {Effect of spatial selection on velocity dispersions. \emph{Left}:
IC stars selected from the model of \citet{as15} at $R=8\pm1\kpc$ and either
$|z|<0.1\kpc$ (solid) or no restriction in z (dashed). We track these stars
back in time and at each output calculate their velocity dispersions in
radial (black) and vertical (red) directions. \emph{Middle and right}: AVRs
for true ages in models YN1 and Y1 in radial (black) and vertical (red)
directions at $R=8\pm0.5\kpc$. Solid lines are for stars at $|z|<100\pc$ and
dashed lines are for all stars irrespective of $z$ position.  }
\label{select}
\end{figure*}

Despite the agreement between degraded AVRs of Model E2 and Snhd data noted
in Section \ref{sec:whichModel}, the model's values for $\beta$ conflict with
the data. Most prominently $\beta_z=1.1$ is significantly higher than
$\beta_z\sim0.45-0.53$ found in the Snhd (\citealp{holmberg}, AB09).  By
contrast, the Y models with GMCs yield values of $\beta$ that are consistent
with the data, even though these models lack a thick disc.  As far as
in-plane heating is concerned, $\beta_R$ is lower in all models than inferred
for the Snhd, where $\beta_R\sim0.31-0.39$.

\subsubsection{Radial variation of AVRs}

Fig.~\ref{heatrad} explores how AVRs are predicted to vary with radius in
four models selected because they show different behaviours.
Each panel in the first three rows shows
eight sets of dots, with each such set showing the AVR given by the true ages
of stars found at a radius in the range $(3,10)\kpc$: the larger the value of
$R$, the lower the curve lies in the panel. Fits of equation
\eqref{eq:heatlaw} to each curve are over-plotted in colours that move
through the spectrum from blue to orange as $R$ increases. The values of $\beta$
for these curves are plotted in the bottom row of panels, with black stars
from $\sigma_R(\tau)$, green crosses from $\sigma_\phi(\tau)$ and red
diamonds from $\sigma_z(\tau)$.  For $R=8\kpc$, we also show how using
degraded rather than true ages changes $\beta$. The in-plane values for
$\beta$ are small but increase outwards. The vertical value of $\beta$ is
much larger and its radial variation is rather various.

Model Y1s2, shown in the second column of Fig.~\ref{heatrad}, is a typical
model with GMCs and a weak bar at $R\lesssim3\kpc$.  Its AVRs are well fitted
by equation \eqref{eq:heatlaw} at all radii.  The in-plane velocity
dispersions for the youngest age bins increase mildly with decreasing radius,
whereas young stars at all radii are equally cold vertically. Outside
$R=3\kpc$ the in-plane values of $\beta$ are almost independent of $R$, while
$\beta_z$ declines gently with increasing $R$.  The shape of the AVR thus
depends only mildly on radius.

As far as vertical heating in the outer disc is concerned, we note that
stars which have migrated outwards have higher velocity dispersions than
stars which have not migrated, as qualitatively predicted by \citet{sb09,sb12}.
Non-migrating stars still show a clear increase of $\sigma_z$ with age
due to local GMC heating.

In contrast to Model Y1s2, Model Y1, shown in the leftmost column of
Fig.~\ref{heatrad}, has a strong buckled bar of length $L_{\rm bar}\sim 4
\kpc$. Its vertical AVRs behave like those of Model Y1s2, but the bar
significantly changes the in-plane AVRs. In particular, at the youngest ages
the velocity dispersion increases rapidly with decreasing $R$ because the
bar's gravitational field deforms orbits from circular ones.  On
account of our use of an adaptive cutoff, no young stars were inserted into
the bar region after $t\sim 7 \gyr$, and the youngest stars at $R=3-4\kpc$
are there because they have been captured onto bar orbits.  The influence of
the bar moves the disc region, in which the shape of the AVR is almost
independent of $R$, outwards to $R\gtrsim6\kpc$.

Model Y4f$\zeta$-, shown in the third column of Fig.~\ref{heatrad}, 
has a thicker disc due to the formation of extended ($R\sim10\kpc$) $m=3$ and
$m=2$ non-axisymmetric structures at $t\sim 6\gyr$.  At  young ages the
in-plane AVRs of this model
are affected even more strongly by global non-axisymmetries. A feature in the
AVR caused by the event at $t\sim 6\gyr$ is visible at $\tau\sim 4\gyr$ at
all displayed radii for the in-plane dispersions. On account of these
features, equation \eqref{eq:heatlaw} provides an unusually poor fit to the
data.

Model E2 in the rightmost column of Fig.~\ref{heatrad} features a thick disc
in its IC and has a long bar ($L_{\rm bar}\sim 6 \kpc$) at $t=10\gyr$. In the
bottom right panel we see the impact on the vertical AVR of the buckling of
this bar, which created an X-shaped region out to $R\sim 4\kpc$. In contrast
to the other models, at the end of the third row of Fig.~\ref{heatrad} we see
that at the youngest ages $\sigma_z$ increases significantly with decreasing
radius. The thick disc from the IC shows up in the AVR as an abrupt increase
in $\sigma(\tau)$ at $\tau=10\gyr$: this increase  is particularly evident for
$\sigma_z(\tau)$ at the largest radii. On account of this step in $\sigma$,
equation \eqref{eq:heatlaw} provides an unusually poor fit to the data.

\section{Heating histories of coeval populations}
\label{sec:heathistory}

We now extract the intrinsic heating histories of coeval populations that
make up the AVR at a certain time and place. To relate this to the Snhd AVR
we select stars at $t=10\gyr$ and $R=8\pm1\kpc$, then divide
them into age groups $\Delta \tau =0.05 \gyr$ wide and track them from their
time of birth $t_{\rm b}$ until $t_{\rm f}=10\gyr$.  In this way, we
calculate velocity dispersions of each group at every output and thus
assemble a heating history for each coeval cohort.

\subsection{Selection effects}

Selecting stars from a limited spatial volume introduces phase correlations
between stars which influence the values of velocity dispersions at the time
of selection and before. For example, when stars are selected close to $z=0$,
they are all close to their maxima in $|v_z|$. Tracking them back in time,
their vertical velocity dispersions thus have to be lower than at the time of
selection. Unfortunately, the output frequency (snapshots at $50\myr$
intervals) of the simulations discussed here is too low demonstrate this
effect clearly. So we illustrate the effect with a simulation from
\citet{as15} that is similar to YN1, but has a lower number of particles in
the dark halo and a gas component. For this model we have snapshots at
$1\myr$ intervals.

\begin{figure*}
\centering
\includegraphics[width=18cm]{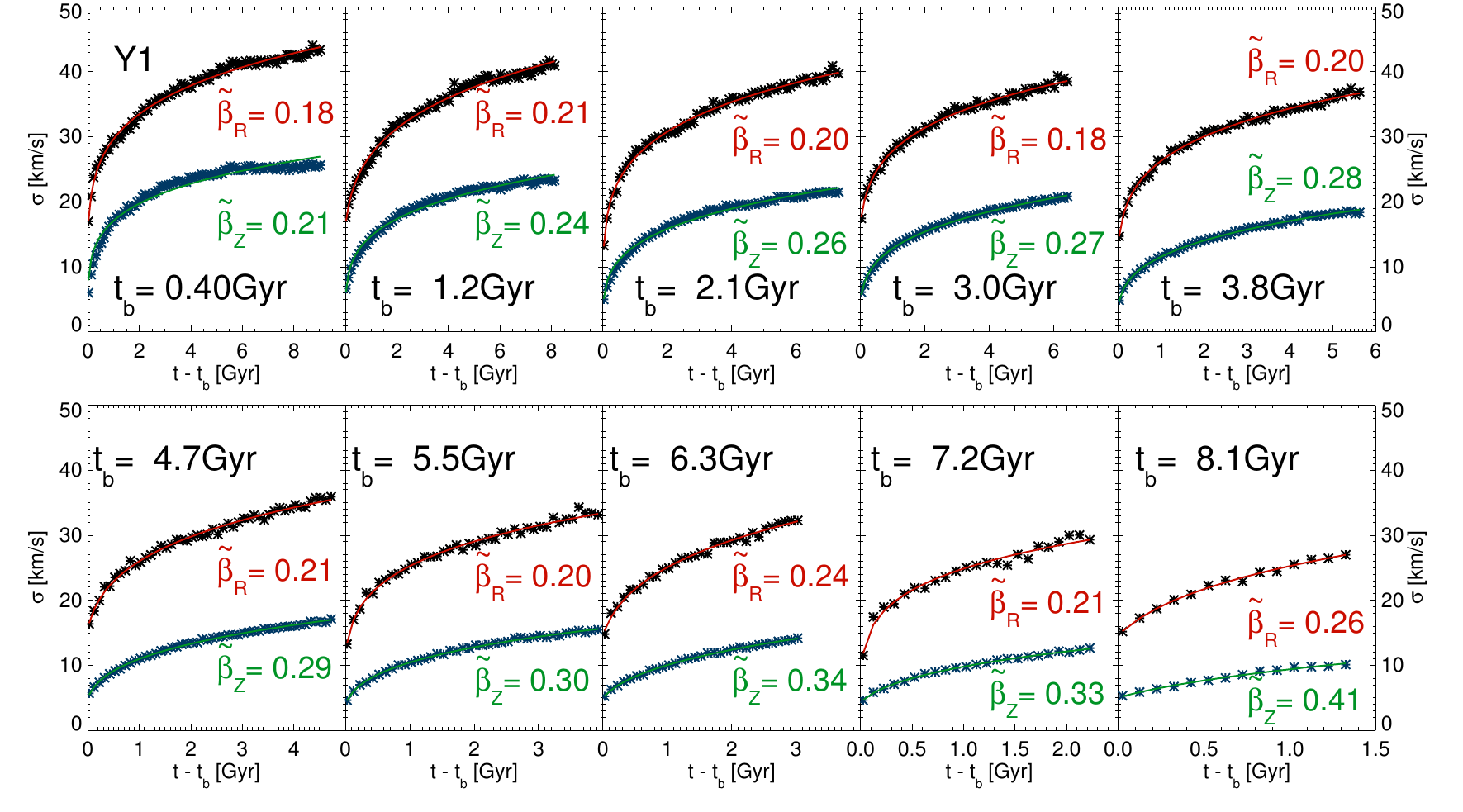}\\
\caption
{The evolution with time $t-t_{\rm b}$ of the vertical and radial velocity 
dispersions in Model Y1 for stars with different birth times $t_{\rm b}$. 
The stars are selected to be at $R=8\pm0.5\kpc$ at $t=10\gyr$.
Displayed are curves for ten different age cohorts. Overplotted are fits of
$\sigma(t)=\tilde{\sigma}_{10}\left[\left(t+\tilde{\tau}_1\right) / \left(10\gyr + \tilde{\tau}_1\right)\right]^{\tilde{\beta}}$. 
The fitted values of $\tilde{\beta}$ are displayed in red ($\tilde{\beta}_R$) and green ($\tilde{\beta}_z$). 
}
\label{allheat}
\end{figure*}

From the \citet{as15} model, we select IC (i.e. old) stars at $t=7.8\gyr$,
$R=8\pm1\kpc$ and $|z|\le0.1\kpc$ and track them back for $200\myr$. From the
full red curve in the
left panel of Fig.~\ref{select} we see that the vertical dispersion
$\sigma_z$ of these stars undergoes an oscillation with a period of $\sim
50\myr$, which is only distinctly visible in the last period before selection.
This period is the average vertical oscillation period of these stars and
phase mixing diminishes the correlation with increasing time before selection
near the midplane.  Repeating the experiment with only radial and no vertical
selection bounds imposed (dashed lines), the effect disappears and $\sigma_z$
is almost constant.

Selecting stars in radius also imposes phase correlations. E.g., if a disc
galaxy has an old, centrally concentrated disc population, old stars at
$R=8\pm1\kpc$ will be dominated by stars on eccentric orbits, which have
their guiding centre radii further in and will therefore be selected close to
apocentre. This effect alone however cannot explain the tracked-back radial 
velocity dispersions (black lines) in the left panel of Fig.~\ref{select}. 
As predicted, they show oscillations of $\sim 10$ per cent, but these 
oscillations appear to have more than one underlying period of $\sim 100\myr$,
irrespective of the $z$ selection of the stars.

On account of these selection effects, we decided to track heating histories
in the following way: we select stars at $t=10\gyr$ and $R=8\pm1\kpc$, but do
not select in $z$.  To prevent oscillations impacting our results, we exclude
the last $500\myr$ from our analysis -- that is, we fit $\sigma(t-t_{\rm b})$ between
$t=t_{\rm b}$ and $t=9.5\gyr$. We fit the curves $\sigma(t-t_{\rm b})$ to equation
\eqref{eq:heatlaw} with $x=t-t_{\rm b}$. To avoid confusion, we mark parameters derived
from heating histories with a tilde, i.e. we write $\tilde{\beta}$,
$\tilde{\sigma}_{10}$ etc.

We compare these parameters to the parameters from AVRs determined at
$t=10\gyr$ and $R=8\pm0.5\kpc$, without a $z$ selection. Fig.~\ref{azi} has
already shown that AVRs depend little on vertical selection range. We
reiterate this point in the middle and right panels of Fig.~\ref{select},
where we show for models YN1 and Y1 radial and vertical AVRs for $|z|<100\pc$
(solid lines) and for all stars irrespective of $z$ position (dashed lines).
Standard models with a single-exponential vertical profile, such as Y1, show
no significant difference between solid and dashed lines. Other models, like
YN1, show only mildly higher dispersions for old stars when considering stars
at all $z$.  The AVR parameters thus differ little between the two ways of
selection of stars for all models. The most notable difference is that
$\beta$ tends to be mildly higher when the selection of stars is unrestricted in $z$.

\subsection{How heating histories shape AVRs}

In Fig.~\ref{allheat} we plot the heating histories of several coeval
populations from Model Y1 that end up at $R=8\kpc$. The extracted data are
shown as black ($\sigma_R$) and blue asterisks ($\sigma_z$), whereas the fits
are overplotted as red ($\sigma_R$) and green lines ($\sigma_z$).  Equation
\eqref{eq:heatlaw} again provides very good fits to the data, regardless of
the population's time of birth. In each panel, we give the fitted values
of $\tilde{\beta}$. We note that $\tilde{\beta}_R$ fluctuates between 0.18 and 0.26, whereas
$\tilde{\beta}_z$ slowly increases with $t_{\rm b}$ from 0.21 to 0.41. Comparing these
values to the ones found for Y1 in the AVRs of Fig.~\ref{heatfits} we note
that $\tilde{\beta}_R$ shows similar values as $\beta_{R}$, whereas $\tilde{\beta}_z$
is always smaller than $\beta_z$, irrespective of the choice of true or degraded ages.

\begin{figure*}
\centering
\includegraphics[width=18cm]{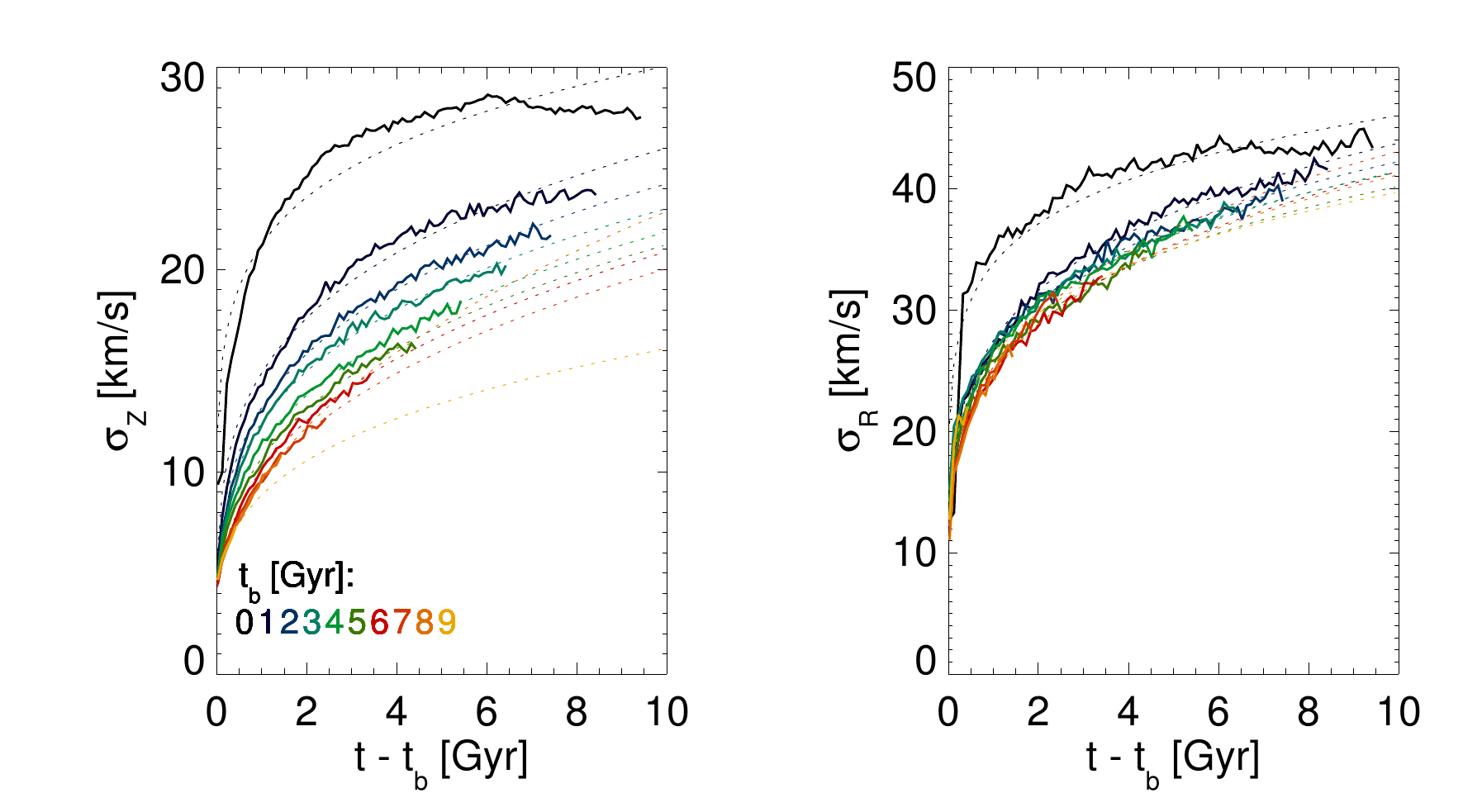}\\
\caption
{The evolution with time $t-t_{\rm b}$ of the vertical (left panel) and radial 
(right panel) velocity dispersions in Model Y1 at $R=8\kpc$ for stars with 
different birth times ranging from IC stars (black) to stars born after $t=8\gyr$
(orange). Overplotted as dotted lines of the same colour are fits of equation 
\eqref{eq:heatlaw} to the curves, which are extended to 10 Gyr.
}
\label{evoheat}
\end{figure*}

In Fig.~\ref{evoheat} we plot the heating curves shown in
Fig.~\ref{allheat} in a different way by showing several curves corresponding
to populations of different birth times (encoded in colour) on top of each other.
The left panel shows $\sigma_z$, whereas the right panel shows $\sigma_R$. 
In dotted lines of corresponding colours we overplot fits of equation 
\eqref{eq:heatlaw} to the curves and extend these fitted
curves to $10\gyr$ to show how the populations would evolve if they continued
to heat according to equation \eqref{eq:heatlaw}.

The curves for $\sigma_z(t-t_{\rm b})$ clearly demonstrate that the heating 
histories vary with time of birth. This has already been indicated by the
increase in $\tilde{\beta}_z$ with $t_{\rm b}$ shown in Fig.~\ref{allheat}. But this
alone would not cause the large differences. The value $\sigma_z(t-t_{\rm b})$
would reach after 10 Gyr, as encoded by the parameter $\tilde{\sigma}_{10}$ from 
equation \eqref{eq:heatlaw}, also decreases strongly with increasing $t_{\rm b}$.
The reason is easily understood.  As ABS16 showed, vertical heating in 
these models is dominated by GMCs, and for standard parameters (e.g., 
Model Y1), the fraction of disc mass residing in GMCs is $\sim 30$ per cent
at early times and steadily decreasing. At $t=10\gyr$, $<5$ per cent of the 
mass remains in GMCs, consistent with observations of the MW.  Consequently,
the global efficiency of GMC heating decreases and the intrinsic heating 
history evolves towards smaller $\tilde{\sigma}_{10}$ with mildly varying $\tilde{\beta}_z$.  

The AVR for any model is given by the lower envelope of the solid heating curves 
in its analogue of Fig.~\ref{evoheat}. A glance at the left panel of 
Fig.~\ref{evoheat} shows that the exponent $\beta_z$ required to fit this
envelope must be higher than the values of $\tilde{\beta}_z$ associated with 
any individual heating history. Consequently the AVR $\sigma_z(\tau)$ is
shaped by the \emph{evolution} of the heating history and there is no 
universal heating history valid for all stellar populations as was assumed
by AB09 and many other authors.

The situation regarding the heating histories $\sigma_R(t-t_{\rm b})$ in the
right panel of Figure \ref{evoheat} is markedly different.  The tendency to
stronger heating of populations born at the earliest times remains visible,
but curves for $t_{\rm b}\ga1\gyr$ lie almost on top of one another, indicating
that after $\sim 1\gyr$ the heating history hardly changes. From this fact
it follows that the difference between the in-plane AVRs and heating 
histories is mild.  In the models of ABS16, after the first Gyr, when the 
high mass fraction of GMCs can lead to powerful radial heating by clumps 
of GMC particles, in-plane heating is dominated by spiral structure and the 
bar. These non-axisymmetries are constantly exited by the addition of cold
stellar populations: in the region of the disc, in which its rotation curve
is not dominated by the contribution from the dark halo, and which lies 
outside of the bar, Toomre's $Q$ \citep{toomre} parameter settles to a 
characteristic value $Q\sim 1.5$. This requires the velocity dispersion
$\sigma_R$ of the entire disc to increase (see Fig. 9 in ABS16). In 
consequence, young stars which are born cold need to be heated efficiently at all
times, which results in almost constant in-plane heating histories.

\subsection{Heating histories in standard models}

\begin{figure*}
\centering
\includegraphics[width=18cm]{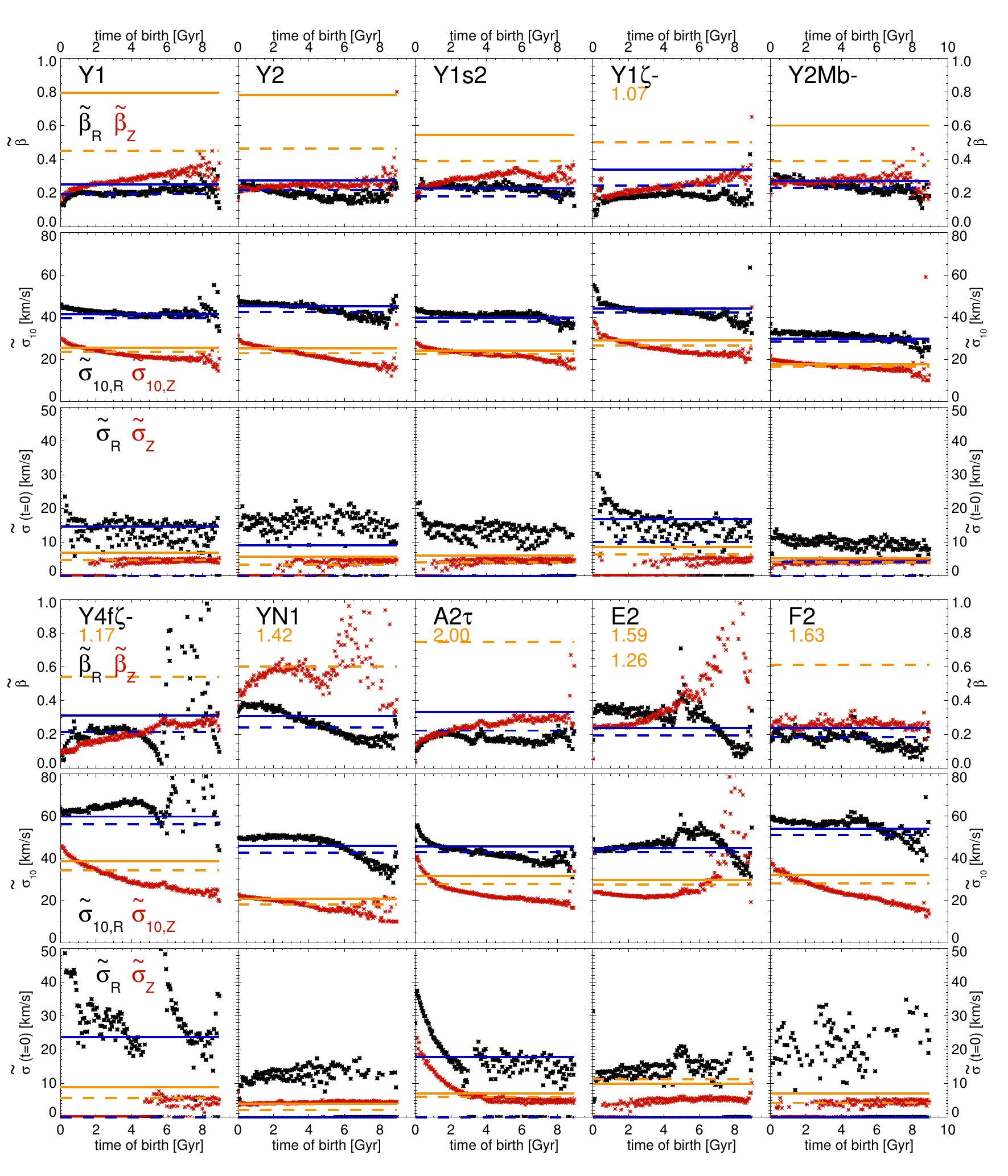}\\
\caption {Parameters from fits of equation \eqref{eq:heatlaw} to heating
histories $\sigma(t-t_{\rm b})$ as functions of time of birth $t_{\rm b}$ of stars found
at $R=8\kpc$ after $t=10\gyr$. Three panels are shown for each model: $\tilde{\beta}$
(upper); $\tilde{\sigma}_{10}$ (middle); $\tilde{\sigma} (t=0)$ (lower) (the latter is
determined by the parameter $\tilde{\tau}_1>0$). Parameter values $\beta$, $\sigma_{10}$
and $\sigma (t=0)$ found by fitting
equation \eqref{eq:heatlaw} to the AVRs are indicated by horizontal lines:
solid lines when true ages are used and dashed lines when degraded ages are
used. When the AVR yields $\beta>1$, the value is printed on the panel's left
side.  } \label{betaE}
\end{figure*}

In Fig.~\ref{betaE} we plot for the heating histories of several models the
parameters $\tilde{\beta}$ (first and fourth rows), $\tilde{\sigma}_{10}$ (second and fifth
rows) and $\tilde{\sigma}(t=0)$ (third and sixth rows) as functions of $t_{\rm b}$
for both vertical (red asterisks) and radial (black asterisks) heating histories
extracted from analogues of Fig.~\ref{allheat}. $\tilde{\sigma}(t=0)$ is determined
by the parameter $\tilde{\tau}_1$ of equation \eqref{eq:heatlaw}, which is required
to be $\tilde{\tau}_1\ge0$, so that $\tilde{\sigma}(t=0)\ge0$.  To display the differences
between heating histories and AVRs at $R=8\kpc$ we show in each panel by
horizontal lines the values of the vertical (in gold) and radial heating
parameters (blue) extracted from the corresponding AVRs 
at $R=8\pm0.5\kpc$ without $z$ restriction: full lines for the
values obtained using true ages and dashed lines for values obtained using
degraded ages.  When $\beta>1$, so the line lies beyond the top of the panel,
the value is printed in the panel. Note that due to the difference in $z$ selection 
the AVR parameters differ mildly from those found in Section \ref{sec:heatIndex}.

The models analysed in the first three rows of Fig.~\ref{betaE} are variations on
the standard Model Y1. They all have single-exponential vertical profiles and
bars of varying extent and strength. We note that the $\tilde{\beta}$ values change
little with the time of birth of the populations.  The analytic treatment of
scattering by GMCs in \citet{lacey} yielded $\tilde{\beta}=0.25$, and \citet{haenninen}
found $\tilde{\beta}_R=0.21$ and $\tilde{\beta}_z=0.26$ from numerical simulations of heating
by GMCs. \citet{desimone} found that $\tilde{\beta}_R$ caused by transient spirals
can lead to a wide range of $\tilde{\beta}_R\sim0.2-0.7$ depending on properties 
of the spirals. These models did not strictly distinguish between AVRs and 
heating histories. As their heating indices were derived from the evolution
of velocity dispersions, we assign tildes. The values for the heating index 
$\tilde{\beta}$ from our simulations agree well with these results, as
$0.15<\tilde{\beta}_R<0.25$ and $0.20<\tilde{\beta}_z<0.33$.

For the oldest populations in all models, we find $\tilde{\beta}_z\approx\tilde{\beta}_R$,
whereas for the younger populations $\tilde{\beta}_z>\tilde{\beta}_R$. Lower $\tilde{\beta}$
indicate a stronger initial increase in velocity dispersions and an earlier saturation of heating.
For the younger populations the explanation is likely that after
in-plane heating, driven by spiral structure, has essentially saturated, GMCs
continue to increase $\sigma_z$ by deflecting stars from eccentric
near-planar orbits to less eccentric and more highly inclined orbits.
For the populations born early on, GMCs have a high mass fraction in the disc and
can efficiently heat the disc both vertically and radially. Consequently, the high 
GMC mass fraction in Y1$\zeta$- leads to $\tilde{\beta}_z\approx\tilde{\beta}_R$
for a wider range of $t_{\rm b}$, whereas in Y2Mb- the lower disc mass leads to less
efficient spiral heating and thus a longer period of GMCs heating the disc
both radially and vertically.

For Models Y1, Y1s2 and Y1$\zeta$-, which grow inside-out, 
there is a mild increase in $\tilde{\beta}_z$ visible with
increasing time of birth whereas for Models Y2 and Y2Mb-, which have a
constant input scalelength $h_R=2.5\kpc$, there is a very mild decrease in
$\tilde{\beta}_R$. It is striking that $\tilde{\beta}_z$ for all these five models and for
the vast majority of times of birth are lower than the values of $\beta_z$
inferred from the AVR.  In contrast, the values of $\tilde{\beta}_R$
scatter around the values $\beta_R$ from AVRs for all models with the exception
of Model Y1$\zeta$-, which has a higher fraction of its mass in GMCs and for which
the value of $\beta_R$ is consistently higher than $\tilde{\beta}_R$ by $\sim
0.05-0.15$ depending on which ages are used.

For all models, the values of $\tilde{\sigma}_{10}$ scatter around the values
$\sigma_{10}$ found from AVRs. Naively one might think the heating history of
the oldest cohort to agree with the AVR at the oldest ages. However, in
several panels of Fig.~\ref{betaE} $\tilde{\sigma}_{10,R}$ for $t_{\rm b}=0$
is larger than expected by this reckoning. This discordance arises, as
Fig.~\ref{evoheat} illustrates, because equation \eqref{eq:heatlaw} often
provides a poor fit to the heating history of the very oldest stars, and
tends to over-estimate their $\sigma$ at large values of $t-t_{\rm b}$: in
reality for these stars $\sigma$ saturates earlier than equation
\eqref{eq:heatlaw} predicts. 

Fits to $\sigma_z(t-t_{\rm b})$ yield values of $\tilde{\sigma}_{10}$ that decrease
with $t_{\rm b}$, just as Fig.~\ref{evoheat} indicates for Model Y1. This
decline is weaker in Y1s2 as its SFR and thus its total GMC mass decline on a
longer timescale.  Consequently, the AVR of this model yields a smaller value
of $\beta_z$ than the AVRs of other models.  As far as radial heating
histories are concerned, Y1$\zeta$- and to a lesser degree Y2 show a more
significant decline with increasing $t_{\rm b}$ in $\tilde{\sigma}_{R,10}$ than Y1, which explains why AVR values
of $\beta_R$ are mildly larger than the $\tilde{\beta}_R$ values in these models. 

$\tilde{\sigma}_{R}(t=0)$ for all models mostly scatters between 10 and $15\kms$, whereas 
$\tilde{\sigma}_{z}(t=0)$ mostly lies between 0 and $6\kms$. We note that the fitted 
$\tilde{\sigma}_{i}(t=0)$ does not necessarily describe the actual $\sigma_{i}$ of very 
young stars well, as a best fit can deviate from the fitted data. Still, this 
finding reiterates that the radial dispersions of stars increase almost 
instantaneously after insertion in reaction to the local non-axisymmetries, 
whereas the vertical dispersions increase more slowly. For some  fits to radial 
AVRs, $\sigma_{R}(t=0)=0$ is favoured and thus differs from
the typical $\tilde{\sigma}_R (t=0)$ values, whereas for others the values from AVRs
and heating histories are similar. We note that for low $t_{\rm b}$ in Model 
Y1$\zeta$-, we find $\tilde{\sigma}_{R}(t=0)>20\kms$ to be higher than average, indicating 
that a high mass fraction of GMCs can significantly influence the radial heating.

\subsection{Heating histories in non-standard models}

In the fourth to sixth rows of Figure \ref{betaE}, we show models that do not
finish with single-exponential vertical profiles. Model F2
has a lower density halo than Model Y2, but
shows similar evolution of the heating exponents. It develops $m=2$
non-axisymmetries that extend to $R\sim 10 \kpc$. These lead to smaller
values $\tilde{\beta}_R\sim 0.1$ for stars born at late times. The long bar
causes high values of $\tilde{\sigma}_{R,10}\sim60\kms$. The values of
$\tilde\beta$ and $\tilde\sigma_{10}$ scatter much less than do the values of
$\tilde{\sigma}_{R}(t=0)$, which cover the range $(0,35)\kms$, presumably
because this parameter has the smallest influence on the fits.  Although we
have seen that the AVRs of Models F2 and Y2 yield significantly different
values of $\beta_z$, the heating histories of these two models yield almost
identical values of $\tilde{\beta}_z$. As was discussed in Section
\ref{sec:heatIndex}, Model F2 has a large bar early on and thus at early
times the cutoff radius and the early surface density of GMCs at $R\sim 8
\kpc$ are both larger than in Y2.  Consequently, the efficiency of GMC
heating and thus $\tilde{\sigma}_{z,10}$ decline more strongly in F2, causing
the fit to the AVR $\sigma_z(\tau)$ to yield a larger value of $\beta_z$.
Direct vertical heating by the bar likely plays a role for this model as
well, but disentangling the effects is beyond the scope of this paper.

Model A2$\tau$ has a thicker disc in its ICs than Model Y2 and an input
velocity dispersion $\sigma_0$ that decreases with time
(eq.~\ref{sigmazerot}), so stars at
early times are born significantly hotter than stars at late times, which are
born as cold as those in Model Y2. The declining input dispersions are nicely
visible in $\tilde{\sigma}(t=0)$. As noted in ABS16, vertical dispersions for younger
stars are lower than input dispersions $\sigma_0$ as all stars are added at $z=0$
and thus lose kinetic energy when moving away from the midplane. Interestingly,
this decline in $\tilde{\sigma}(t=0)$ is also reflected in a milder decline in $\tilde{\sigma}_{10}$
and smaller values of $\tilde{\beta}$ in both vertical and radial directions
for stars born at early times. Small values of $\tilde{\beta}$ are likely connected to an earlier
saturation of heating due to higher initial dispersions. At late times we find only
mild differences between the heating indices of Model A2$\tau$ and Model Y2.

Model E2 has an even thicker and more extended disc in its IC than Model
A2$\tau$, and the same small, constant input dispersion $\sigma_0$ as Model
Y2.  At early times the thick, extended disc suppresses spiral and bar
formation below the level seen in the thin, compact disc in the IC of Model
Y2. Consequently, early on radial heating is less powerful in Model E2, with
the consequence that the oldest populations heat more slowly in Model E2 and
thus have large $\tilde{\beta}_R>0.3$ as the saturation phase happens later (fourth
panel in second row). A long ($L_{\rm bar}\sim 6\kpc$), strong and buckled
bar forms in E2 after $t\sim 6\gyr$. This bar significantly influences the
heating of the populations which end up at $R=8\kpc$. For populations born
after bar formation, $\tilde{\beta}_R$ decreases strongly as a strong bar leads to a
very fast increase in $\sigma_R$ for young stars and thus a quicker
saturation and a low value of $\tilde{\beta}_R$. At the same time $\tilde{\sigma}_{R,10}$
decreases, implying that populations born at late times are expected to
attain lower velocity dispersions after $10\gyr$. This is likely driven by
the lower $\tilde{\beta}_R$, as for these stars there is no information
available at
$t\gtrsim4\gyr$ and thus the fits likely over-predict the saturation effect
and thus under-predict the continuous in-plane heating. By adding an
additional vertical heating mechanism for young stars, bar buckling increases
$\tilde{\beta}_z$ and $\tilde{\sigma}_{z,10}$. Then the fits very likely over-predict the
vertical dispersion these stars would attain after $10\gyr$.

A different situation is found in the leftmost panel of the second row for
Model Y4f$\zeta$-. In this model, which has a very compact feeding history, a
fixed cutoff and a high GMC mass fraction, strong and extended $m=3$ and
$m=2$ modes form around $t\sim 6\gyr$. These events lead to deviations
from simple $t^{\beta}$ AVRs and heating histories as was shown in ABS16
and in Fig.~\ref{heatrad}. Consequently the quality of fits of equation 
\eqref{eq:heatlaw} to the heating histories is worse than for other models
and there are strong variations in the radial heating parameters. $\tilde{\beta}_R$
varies between 0 and 1 for stars born after $5\gyr$, $\tilde{\sigma}_{R, 10}$ is generally
high at $\sim 60\kms$ due to the strong non-axisymmetries and also scatters strongly
for stars born after $5\gyr$, and $\tilde{\sigma}_{R}(t=0)$ varies systematically between
0 and $50\kms$. The decrease in $\tilde{\sigma}_{z,10}$ is strong, which explains the 
high AVR value of $\beta_z$. 

Finally, we also show one model without GMCs, Model YN1 (second panel second
row). Both the radial and vertical heating histories have $\tilde{\beta}$ values that
differ sharply from those of the standard Y models as the heating curves are 
shaped only by non-axisymmetric structure. $\tilde{\beta}_R$ decreases from
0.4 to 0.15 with increasing time of birth, whereas $\tilde{\beta}_z$ is rather high,
with typical values $\tilde{\beta}_z\sim 0.5-0.6$ and a tendency to increase with
$t_{\rm b}$. As in model E2, the decrease in $\tilde{\beta}_R$ and the decrease
in $\tilde{\sigma}_{R, 10}$ for stars born at late times, is likely caused by a
strong bar, which in YN1 forms early and extends to $R\sim5\kpc$ by
the end of the simulation. On account of the change in the main source of
vertical heat, the value of $\tilde{\beta}_z$ for Model YN1 
differs sharply from the values yielded by models with GMCs.

\section{Discussion}
\label{sec:discuss}

In a galactic disc the random velocities of stars are increased by the
fluctuating non-axisymmetric component of the gravitational field. GMCs and
spiral arms are both major contributors to this component. 

In a naive picture of the heating of the solar neighbourhood, stars diffuse
through velocity space from the velocities of circular orbits under the
influence of fluctuations that constitute a  stationary  random process
\citep{wielen}. If the diffusion coefficient that governed this process were
independent of ${\bf v}$, the velocity dispersion of a coeval cohort of stars
would grow as $(t-t_{\rm b})^{1/2}$. In reality the diffusion coefficient
must be a declining function of $|{\bf v}|$ because a given potential
fluctuation deflects fast stars through smaller angles than slow stars
(typically $\delta\theta\propto v^{-2}$). Hence
if the fluctuating component of the gravitational potential is statistically
stationary, the exponent in the {\it heating law} $\sigma\propto(t-t_{\rm
b})^{\tilde{\beta}}$ has to be less than $0.5$: \citet{lacey} found $\tilde{\beta}=0.25$ for
analytical models of GMC heating. In the case of spiral structure the concept
of a deflection angle is problematic, but stars with significant random
velocities and therefore eccentric orbits, encounter a given spiral wave at a
variety of orbital phases, so the time-averaged impact of the wave on a fast
star is small. That is, very general physical principles leave no doubt that
in the presence of statistically stationary fluctuations, the exponent in any
heating law is $\la0.25$ \citep{binneyNAR}.

The black and red asterisks in the first and forth rows of 
Fig.~\ref{betaE} plot the exponents of heating
histories rather than heating laws: they relate to the variation with
$t-t_{\rm b}$ of $\sigma_R$ and $\sigma_z$ for a group of stars that are now
at $R=8\kpc$ rather than the variation with $t-t_{\rm b}$ of a group of stars
that formed at $R=8\kpc$. Nonetheless, these two groups are similar, so it is
interesting to see that in the standard Y models $\tilde{\beta}_R$ tends to be
constant at a value $\la0.25$, consistent with fairly stationary
fluctuations. In some Y models $\tilde{\beta}_z$ is constant at a similar or slightly
larger value than $\tilde{\beta}_R$, and in other models it increases mildly 
with $t_{\rm b}$, remaining at $\tilde{\beta}_z<0.33$ for most coeval cohorts.

Spiral structure in discs with $Q>1$ is maintained by swing amplification of
noise \citep{toomre2,fouvry}. Since the efficacy of swing amplification rises
very sharply as Toomre's
\begin{equation}
Q\equiv{\sigma_R\kappa\over3.36G\Sigma}
\end{equation}
 drops towards unity, a cold ($Q\la1$) stellar disc very rapidly heats until
$Q\simeq1.5$. Hence from an early time, a galactic disc will have
$Q\simeq1.5$. As the disc grows in mass, its surface density $\Sigma$ rises
and $\sigma_R$ is driven upwards by spiral structure to keep $Q$
approximately constant (see Fig. 9 in ABS16). Hence, the time dependence of 
$\sigma_R$ for stars of all ages at a given radius is essentially 
prescribed by the history of gas accretion. As a consequence, young stars 
which are born cold are heated efficiently by constantly excited
spiral structure at all times.

Spiral structure heats within the plane but does not directly increase
$\sigma_z$. GMCs, by contrast heat the disc vertically as well as
horizontally. Consequently, the ratio $\sigma_z/\sigma_R$ is a measure of
the relative importance of GMCs and spirals as heating agents.

Fig.~\ref{betaE} reveals that for almost all heating histories,
$\tilde{\beta}_z>\tilde{\beta}_R$. Consequently, the velocity ellipsoids of groups of coeval
stars tend to become rounder over time. The natural explanation is that after
in-plane heating, driven by spiral structure, has essentially saturated, GMCs
continue to increase $\sigma_z$ by deflecting stars from eccentric
near-planar orbits to less eccentric and more highly inclined orbits.

One may plausibly argue, as ABS16 did, that the mass of gas in all GMCs is
proportional to the SFR, since over the lifetime of a GMC a fraction $\zeta$
of the GMC's mass is converted to stars. Hence, the rate at which GMCs heat
the disc declines with the SFR. Moreover, early on the disc has a low mass, 
so each GMC represents a larger fraction of the total disc mass. Since the 
rate at which an individual GMC heats scales with its mass relative to the 
mass of the Galaxy interior to its orbit, each GMC is individually a more 
effective heating agent early in the life of the disc. Hence over time the 
heating power of the ensemble of GMCs declines faster than the SFR. Large-scale
spiral structure, by contrast, is associated with the self-gravity of the
stellar disc, which grows steadily over time from a small initial value. That
is, over the life of the disc, the impact of spiral structure has increased
relative to that of GMCs.

If the fluctuations that heat the disc were a stationary random process and
the disc were homogeneous, heating histories would be independent of $t_{\rm
b}$ and have the same functional form as the AVR. We have seen that in the
models heating histories do depend on $t_{\rm b}$ and are very different from
the AVR.  While the dependence of heating histories on $t_{\rm b}$ and their
deviation from the AVR could be entirely attributed to the non-stationary
nature of the fluctuations, a contributing factor is undoubtedly radial
migration \citep{sellwoodb,sb09}, which adds significant complexity to the
problem by mixing stars that were born at small and large radii. Motivated by
the desire to understand data for the Snhd, we have studied samples of stars
that are currently at $R\simeq R_0$.  In a future paper we will consider
groups of stars with a common birth radius. 

\section{Conclusions}
\label{sec:conclude}

In this paper, we have used a series of $N$-body simulations of growing disc
galaxies (ABS16) to study (i) age-velocity-dispersion relations (AVRs)
and (ii) the heating histories of the coeval cohorts of stars which make up the AVRs.
As these models feature heavy GMC particles, secular heating is dominated by a 
combination of scattering of stars off GMCs and non-axisymmetric disc structures. 

To be able to compare these simulations to observational data from the Snhd, 
we analysed the impact on the AVR of biases and errors in
measured stellar ages. Stars with ages $\tau\sim2\gyr$ are very much
over-represented in the GCS data (Fig.~\ref{gcs}). Scattering of such stars to 
young ages artificially boosts $\sigma(\tau)$ at the youngest ages, and depresses
$\sigma(\tau)$ at the oldest ages. When a power law in $\tau$ is fitted to
the measured AVR, lower values of the exponent $\beta$ are recovered than would be in
the absence of errors (see also \citealp{martig}). The reduction in $\beta$ 
is particularly marked in the case of $\sigma_z$ (Fig.~\ref{heatfits}).

On account of spiral structure and bars, AVRs vary with azimuth as well as
radius. Fig.~\ref{azi} quantifies the extent of this azimuthal variation,
which must be borne in mind when considering whether a given model is
consistent with data for the Snhd, which are measured at one particular
azimuth. After taking azimuthal variation into account, we concluded that the
GCS data are consistent with some models in ABS16 that have the expected disc mass
($5\times10^{10}\msun$) and the cosmologically motivated dark halo
($M=10^{12}\msun$, $a=30.2\kpc$). Models with a significantly different disc
mass or a less concentrated dark halo are inconsistent with data for the
Snhd. The data also favour the model that starts with a massive, extended
thick disc over models in which (a rather inadequate) thick disc forms
as a consequence of powerful non-axisymmetries developing in the thin disc.
As we do not self-consistently form appropriate thick discs and as
we lack heating by dark matter substructure, which may contribute a minor
part of the observed disc heating, we are not able to put tight constraints
on our model parameters.

AVRs vary with radius. At locations currently inside the bar, the AVR's index
$\beta_R$ is generally very small, $\beta_R<0.1$, as there are no circular orbits
and young stars thus acquire high $\sigma_R$ rapidly. At the end of the bar,
$\beta_R$ rises abruptly and is thereafter constant or slowly rising with $R$
(Fig.~\ref{heatrad}). By contrast, $\beta_z$ sometimes increases and
sometimes decreases at the end of the bar. A buckling bar can lead to 
exceptionally high $\sigma_z$ for young stars in the bar regions.

The heating history, $\sigma(t-t_{\rm b})$, of stars now in the Snhd that were
born at time $t_{\rm b}$ can also be fitted by the power-law \eqref{eq:heatlaw}.
We mark the corresponding parameters with a tilde.
For standard models, the heating history depends on $t_{\rm b}$ more strongly 
in the case of $\sigma_z$ than $\sigma_R$. Smaller values of the exponent $\tilde{\beta}$ are
required to fit heating histories than AVRs. In fact, values of $\tilde{\beta}_R$ are
consistent with the predictions of dynamics in the case that the fluctuating
gravitational potential is a stationary random process. The values of
$\tilde{\beta}_z$ are generally somewhat larger than is consistent
with a stationary random process, but in agreement with numerical simulations
of stationary GMC heating \citep{haenninen}.

The AVR reflects the history of star and bar formation. The past SFR strongly
affects the AVR for two reasons: the time integral of the SFR
determines the mass of the disc, and thus the fraction of the gravitational
force on a star that derives from the disc rather than the dark halo. At
early times this fraction is small, so spiral arm formation is already suppressed by
a low value of $\sigma_R$. As the mass of the disc increases, spiral
structure increases $\sigma_R$ to keep  Toomre's $Q$ nearly constant. If the
SFR is rapidly declining, the rate at which $\sigma_R$  increases will
decline rapidly, and a relatively small value of $\tilde{\beta}_R$ will be required
to fit the heating histories of the oldest stellar groups.

In contrast, the vertical heating is dominated by GMCs. By analysing
histories of stars that end up at $R=8\kpc$, we showed that the heating
histories of older stars reach higher $\sigma_z$ after $10\gyr$
($\tilde{\sigma}_{10}$) than those of younger stars. The corresponding
$\tilde{\beta}_z$ values only vary mildly with $t_{\rm b}$. This decline in heating
efficiency is connected to the declining influence of GMCs, the total mass of
which declines due to a declining SFR and the mass fraction of which declines
due to a growing stellar disc.  When coeval cohorts, whose
$\tilde{\sigma}_{10}$ values
decline with $t_{\rm b}$, are superposed to form an AVR, a value of
$\beta_z$ in excess of $0.5$ is needed to fit the curve.

By combining all these results we have been able to clarify long standing
discrepancies between the observed AVR and theoretical predictions for
combined spiral and GMC heating. Some of our models correctly reproduce the
general shape of both $\sigma_R(\tau)$ \emph{and} $\sigma_z(\tau)$ as
observed in the Snhd and thus also the ratio of the two components. The key
ingredient is that each coeval cohort of stars that contributes to the AVR
has undergone a different heating history and the AVR is not produced by a
single stationary heating law. We conclude that combined GMC and spiral/bar
heating has likely shaped the MW thin disc AVR.

\section*{Acknowledgements}
We thank the referee for comments that helped improve the paper.
This work was supported by the UK Science and Technology Facilities Council (STFC)
through grant ST/K00106X/1 and by the European Research Council under the European 
Union's Seventh Framework Programme (FP7/2007-2013)/ERC grant agreement no.~321067.
This work used the following compute clusters of the STFC DiRAC HPC Facility 
(www.dirac.ac.uk): i) The COSMA Data Centric system at Durham University, operated
by the Institute for Computational Cosmology. This equipment was funded by a BIS 
National E-infrastructure capital grant ST/K00042X/1, STFC capital grant 
ST/K00087X/1, DiRAC Operations grant ST/K003267/1 and Durham University. 
ii) The DiRAC Complexity system, operated by the University of Leicester 
IT Services. This equipment is funded by BIS National E-Infrastructure capital 
grant ST/K000373/1 and STFC DiRAC Operations grant ST/K0003259/1.
iii) The Oxford University Berg Cluster jointly funded by STFC, the Large 
Facilities Capital Fund of BIS and the University of Oxford. 
DiRAC is part of the National E-Infrastructure.

\label{lastpage}
\end{document}